\documentclass[a4paper,12pt]{article}
\usepackage{epsf,latexsym,amsmath}
\usepackage{times}
\usepackage{amssymb}
\usepackage{graphicx}
\usepackage[all,poly]{xy}
\usepackage{rotating}
\input{epsf}
\xyoption{rotate}
\xyoption{line}
\begin{document}

%\begin{flushright}
%today
%\end{flushright}
%\vspace{24pt}

\begin{center}
{ \Large \bf  $3nj$ morphogenesis and semiclassical disentangling} 
\end{center}
\vspace{24pt}

\begin{center}
{\sl Roger W. Anderson}\\
Department of Chemistry,
University of California, Santa Cruz,\\
 California 95064, USA\\ 

\vspace{12pt}

{\sl Vincenzo Aquilanti}\\
Dipartimento di Chimica,
Universit\`a degli Studi di Perugia\\
via Elce di Sotto 8, 06126 Perugia (Italy)\\ 

\vspace{12pt}

{\sl Annalisa Marzuoli}\\
Dipartimento di Fisica Nucleare e Teorica,
Universit\`a degli Studi di Pavia\\
 and INFN, Sezione di Pavia, 
via A. Bassi 6, 27100 Pavia (Italy);\\ 
E-mail: annalisa.marzuoli@pv.infn.it 
\end{center}

\vspace{24pt}
  
\noindent {\bf Abstract}\\
Recoupling coefficients
($3nj$ symbols) are unitary transformations between 
binary coupled eigenstates of $N=(n+1)$ 
mutually commuting $SU(2)$ angular momentum
operators. They have been used in a variety of applications in
spectroscopy, quantum chemistry and nuclear physics and quite recently also in
quantum gravity  and quantum computing.
These coefficients, naturally associated to
cubic Yutsis graphs, share a number of intriguing
combinatorial, algebraic,  and analytical features
that make them fashinating objects to be studied on their own. 
In this paper we develop a  bottom--up, systematic procedure  for the generation of $3nj$
from $3(n-1)j$ diagrams  by resorting to  diagrammatical and 
algebraic methods. We provide also a novel approach to the problem of 
classifying various regimes of semiclassical expansions of
$3nj$ coefficients (asymptotic disentangling of $3nj$ diagrams) 
for $n \geq 3$
by means of combinatorial, analytical and numerical tools.

\vspace{12pt}

\


\vfill
\newpage

\section*{Introduction}

In the quantum theory of angular momentum --mathematically encoded in $SU(2)$ 
representation theory-- the study of (re)coupling of pure eigenstates of several,
mutually commuting angular momentum operators $\mathbf{J}_1,\mathbf{J}_2,
\dots \mathbf{J}_N$ is considered an advanced topic, and as such it is
not widely known. As pointed out in the classic reference \cite{BiLo9} (topic 12)
\begin{quote}
We shall deal with what is appropriately called {\em the theory
of binary coupling of angular momenta}, since it is a theory in which
one couples angular momenta sequentially, in pairs. (...)
Accordingly, the theory deals with the relationships between different
{\em coupling schemes}  --that is between distinct sequences of pairwise couplings.
This is the content of recoupling theory.\\
The subject is a difficult one and the literature is extensive.
Several monographs ({\em e.g.} \cite{YuLeVa,YuBa,Zar,Russi}\footnote{The last
two monographs were published seven  years after Biedenharn--Louck text;
in particular the last one is now
the most exhaustive and reliable collection of definitions
of angular momentum functions and related algebraic and analytic formulas.})
deal with techniques for implementing "graphical" methods (...).
\end{quote}

This paper is much in the spirit of the program stated soon after that, namely
\begin{quote}
Our goal is to relate coupling methods  to standard results from {\em graph theory}.
The subject divides quite naturally into three parts:\\
(A) (the classification of coupling schemes)\\
(B) the elementary operations underlying the structure of 
transformation (recoupling)
coefficients, or $3nj$ symbols\\
(C) the classification of recoupling coefficients and its relationship 
to the theory of {\em cubic graphs}. 
\end{quote}

We are going to deal first with (B) and (C) above by
formalizing  an effective, step--by--step method  to build up
$3nj$ coefficients of type I, II from $3(n-1)j$s of the same type
(note that, by convention, the integer $n$ is
related to the $N$ angular momenta introduced above by $(n+1)=N$).

Our second  goal  is to take advantage of such a recursive
construction in  the search for a novel unifying
scheme for  addressing semiclassical limits
(asymptotic expansions) of the $3nj$ themselves.
This task calls  into play different techniques, both analytical
and numerical, and is very important in view of practical
applications (that have been discussed extensively elsewhere).

%%%%%%%%%%%%%%%%%%%%%%%%%%%%%%%%%%%%
%%%%%%%%%%%%%%%%%%%%%%%%%%%%                  SECTION 1
%%%%%%%%%%%%%%%%%%%%%%%%%

\section{Recursion method for $3nj$ coefficients:\\ 
general description}

The diagram of any $SU(2)$ $3nj$ coefficient is a cubic
(regular trivalent) graph on $3n$ lines and $2n$ nodes, named Yutsis graph from the
first author of \cite{YuLeVa}. The three edges
stemming from each node are associated to a triad of angular momentum
quantum numbers satisfying triangle inequalities, {\em i.e.} to a
Wigner $3j$ coefficient. 
This property of diagrams reflects the fact that
any $3nj$ can be always written as a sum over all magnetic quantum numbers of
the product of $2n$ (suitably chosen) $3j$ coefficients (see {\em e.g.}
\cite{Russi}, Ch.9).\\
The recursion method for the generation of (all)
$3nj$ diagrams with increasing $n$ may be summarized as in \cite{YuLeVa} p.65
\begin{quote}
({\bf Recursion rule})
The diagrams of the $3nj$ coefficients are obtained 
from the diagrams of the $3(n-1)j$ by inserting two additional nodes 
on any two lines of the diagram
of the $3(n-1)j$ coefficients and joining them together.
\end{quote}
There are some caveats underlying this
combinatorial construction: it must generate diagrams that are not separable 
on fewer than four lines and this is achieved by
requiring that the initial $3nj$ diagram is separable on no fewer than three lines
(examples are provided below); moreover, isomorphic configurations
should be recognized and ruled out 
\footnote{As an aside remark recall that the isomorphism problem in 
graph theory is a typical  
{\bf NP}--complete problem, actually a {\bf \#P} problem,
 even for regular graphs:
this means that, given two instances of graphs, one can decide 
if they are isomorphic or not
in polynomial time but the general solution --the enumeration 
of all graphs isomorphic to a given
one-- requires an exponential time as the complexity of the 
graphs grows, {\em i.e.} as the number
of vertices increases (see {\em e.g.} \cite{GaJo}).}.

Algebraic  expressions ("recurrence formulas") that encode the above combinatorial
prescription on diagrams are provided for the $3nj$ coefficients of the first (I) 
and second (II) type in \cite{YuBa} ((24.8) p.81 and (A.6.14) 
p.143), respectively  and read
\begin{gather}\label{RFtypeI} 
(-1)^{\Phi} \sum_{x} \;(2x +1)\;
 \begin{Bmatrix}
 j_1 & \, & \dots & \, & j_{n-1} & \, & \, &\,\\
 \, & l_1 & \dots  & l_{n-2} & \, & x \\
 k_1 & \, & \dots  & \, & k_{n-1} & \, &\,  &\,
 \end{Bmatrix}
\\ \nonumber
 \begin{Bmatrix}
 j_1 & k_{n-1} & x \\
  l_{n-1} & l_n & k_n
 \end{Bmatrix}\;
 \begin{Bmatrix}
 k_1 & j_{n-1} & x \\
  l_{n-1} & l_n & j_n
 \end{Bmatrix}\;=\;
 \begin{Bmatrix}
 j_1 & \, & \dots & \, & j_n & \,\\
 \, & l_1 & \dots  & \, & \, & \l_n \\
 k_1 & \, & \dots  & \, & k_n &\,
 \end{Bmatrix},
\end{gather}
where $\Phi =j_1-k_1-j_{n-1}+k_{n-1}$, and

\begin{gather}\label{RFtypeII} 
(-1)^{\Psi} \sum_{x} \;(2x +1)\;
 \begin{bmatrix}
 j_1 & \, & \dots & \, & j_{n-1} & \, & \, &\,\\
 \, & l_1 & \dots  & l_{n-2} & \, & x \\
 k_1 & \, & \dots  & \, & k_{n-1} & \, &\,  &\,
 \end{bmatrix}
\\ \nonumber
 \begin{Bmatrix}
 k_1 & k_{n-1} & x \\
  l_{n-1} & l_n & k_n
 \end{Bmatrix}\;
 \begin{Bmatrix}
 j_1 & j_{n-1} & x \\
  l_{n-1} & l_n & j_n
 \end{Bmatrix}\;=\;
 \begin{bmatrix}
 j_1 & \, & \dots & \, & j_n & \,\\
 \, & l_1 & \dots  & \, & \, & \l_n \\
 k_1 & \, & \dots  & \, & k_n &\,
 \end{bmatrix},
\end{gather}
where $\Psi =j_1-k_1+j_{n-1}-k_{n-1}$. Here the angular momenta variables
$\{k_1, \dots, l_1,$ $\dots,  j_1, \dots \}$ run over either
$\{0, 1, 2, \dots \}$ or
$\{\tfrac{1}{2}, \tfrac{3}{2}, \dots \}$ 
and satisfy suitable triangle inequalities. 
The summation variable $x$ is also constrained
by triangle relations.

\vskip 20pt

The roles played by the expressions (\ref{RFtypeI}) and  (\ref{RFtypeII}) 
are multiple and interconnected
\begin{itemize}
\item[i)] at facevalue they  provide
a straightforward  decomposition of a $3nj$ coefficient into a single sum 
involving one $3(n-1)j$ of the same type
and exacly two Wigner $6j$ symbols.\\
(Note however that these decompositions are not  the basic ones 
for computational purposes, since it is well known that both types of coefficients
may be expressed as single sums of products of $n$ $6j$  
symbols \cite{Russi}, Eqns. 1 and 2 p. 361.) 
\item[ii)] from the structural point of view they encode also 
generalizations of the basic
Biedenharn--Elliott identity (see the case of $9j$ generation  in section
2 where these two aspects of  (\ref{RFtypeI}) and  (\ref{RFtypeII}) 
are worked out explicitly).
\item[iii)] taking advantage of symmetries of (any kind of) $3nj$ coefficients
one can equate two different decompositions related by a symmetry 
of a same $3nj$ to get  relations
for the associated $3(n-1)j$ coefficients. For instance the relations 
derived from (\ref{RFtypeI})
have the following general structure (dropping phases and omitting 
all the arguments not relevant
in the present discussion)
\begin{gather}\label{3njsymm}
\sum_x \,
\begin{Bmatrix}
 \bullet  & \bullet  & x \\
  \bullet  & \lambda & \bullet 
 \end{Bmatrix}
\;
\begin{Bmatrix}
 \bullet  & \cdots & \, &\,\\
 \, & \bullet  & \cdots  & x \\
 \bullet  &  \cdots  & \, 
 \end{Bmatrix}
 \;
 \begin{Bmatrix}
 \bullet  &\bullet & x \\
 \bullet  & \lambda  & \bullet 
 \end{Bmatrix}\\ \nonumber 
=\;\sum_y \,
\begin{Bmatrix}
 \bullet  & \bullet  & y \\
  \bullet  & \lambda & \bullet 
 \end{Bmatrix}\;
\begin{Bmatrix}
 \bullet  & \cdots & \, &\,\\
 \, & \bullet  & \cdots  & y  \\
 \bullet  &  \cdots  & \, 
 \end{Bmatrix}
\;
 \begin{Bmatrix}
 \bullet  &\bullet & y \\
 \bullet  & \lambda  & \bullet 
 \end{Bmatrix}.
\end{gather}
Specifying $\lambda=\tfrac{1}{2}, 1,\tfrac{3}{2}, 2,$ {\em etc.} and using explicit 
forms of the resulting $6j$
 symbols, various recursion relations may be obtained (note that the sums on both sides of
(\ref{3njsymm}) reduce to a same number of terms, {\em e.g.} for $\lambda=\tfrac{1}{2}$ one 
would get
$x,y =0,1$). 
 The case of the $9j$ symbol and associated 5--term recursion relation will be worked out 
in section 2.2.

The case of the $6j$ symbol is quite subtle since the well known 3--term 
recursion relation
in one variable
-- used in \cite{PoRe,ScGo}, \cite{BiLo9},  topic 9)
as a second order difference
equation to be analyzed within the WKB framework-- actually derives directly from
(\ref{RFtypeII}) which is nothing but the standard Biedenharn--Elliott identity
(see also (\ref{9jII}) in the next section). This remark has to do with a
deep property of hypergeometric polynomials in  the Askey scheme \cite{Askey}: 
the defining  difference
or differential equation and 
the recursion relation are usually "dual" to each other but the $6j$ happens to 
be "self--dual" so that
the two viewpoints can be used equivalently.
\item[iv)] Decompositions like those given in (\ref{RFtypeI}) and 
(\ref{RFtypeII}), as well as associated recurrence relations, can be also
employed in the opposite direction, namely to show how a $3nj$ symbol
falls back into a $3(n-1)j$ when one of its entry is set to zero. This
well known feature \cite{Russi} will be used in remark (i)
at the end of section 2.2.
\end{itemize}

The bottom--up recursion method, fully developed in this paper by resorting to both 
diagrammatical 
and algebraic  tools \cite{YuLeVa,YuBa,Russi},
represents not only a systematic procedure for the generation of $3nj$'s
from $3(n-1)j$ diagrams (section 3), 
but will provide also a novel approach to the classification
of the (various regimes of) asymptotic disentangling of the coefficients 
themselves (section 4), improving and extending previous results for the
$9j$ found in \cite{AnAqFe}.\\
To make the reader familiar with 
diagrammatic and algebraic tools, in the next section 
we are going to address the case study given
by the $9j$ (I,II).

%\vfill
%\newpage
%%%%%%%%%%%%%%%%%%%%%%%%
%%%%%%%%%%%%%%%%%%%%%%     SECTION 2
%%%%%%%%%%%%%%%%%%%%%%%%%%%%

\section{Genesis of $9j$ coefficients and associated diagrams}

The $9j$(I) coefficient (the "standard" $9j$) can be written as a single 
sum of  three $6j$ symbols (notations as in
\cite{Russi}, Ch. 10 and App. 12).
Note that the following expression can be looked at as the first instance
of the recursion 
formula (\ref{RFtypeI}), namely 
this coefficient arises
as a $3nj$(I) for $n=3$ starting from the unique $3(n-1)j$ with $n=2$, {\em i.e.}
from  the 
$6j$ symbol itself
\begin{equation}\label{9jI}
\sum_x (-)^{2x}(2x+1)
 \begin{Bmatrix}
 a  & b & x \\
 c  & d  & p 
 \end{Bmatrix} 
 \begin{Bmatrix}
 c  & d & x \\
 e  & f  & q
 \end{Bmatrix} 
 \begin{Bmatrix}
 e  & f & x \\
 a & b  & r 
 \end{Bmatrix}\;=\; 
 \begin{Bmatrix}
 a  & f & r \\
 d  & q  & e \\
p & c & b 
 \end{Bmatrix}, 
\end{equation}
where $\text{max} \{ |a-b|,|f-e|,|c-d|\}
\leq x \leq \text{min} \{ a+b,f+e,c+d\}$.\\
The associated Yutsis diagram is 
\[
\xy
0*{\bullet}="A";
<-1cm,1.7cm>*{\bullet}="B";
<0cm,3.4cm>*{\bullet}="C";
<2cm,3.4cm>*{\bullet}="D";
<3cm,1.7cm>*{\bullet}="E";
<2cm,0cm>*{\bullet}="F";
"A";"B" **@{-} ?(0.5)*!/_2mm/{e};
"B";"C" **@{-} ?(0.5)*!/_2mm/{d};
"C";"D" **@{-} ?(0.5)*!/_2mm/{a};
"D";"E" **@{-} ?(0.5)*!/_2mm/{f};
"E";"F" **@{-} ?(0.5)*!/_2mm/{c};
"F";"A" **@{-} ?(0.5)*!/_2mm/{b};
"A";"D" **@{-} ?(0.3)*!/_2mm/{r};
"C";"F" **@{-} ?(0.3)*!/_2mm/{p};
"B";"E" **@{-} ?(0.7)*!/_2mm/{q};
\endxy
\]
from which the triad structure as well as the symmetries of the coefficient are easily 
inferred
(the six triads are associated with 3--valent nodes and correspond 
to  columns and rows of the array; 
symmetries,
up to phases, are implemented by  odd or even permutations of columns or rows and
the value of the coefficient does not change under transposition).

On applying type II recurrence formula (\ref{RFtypeII}) 
one should get a $9j$(II) coefficient
but it can be easily shown that the resulting
configuration is a "trivial" one, being the product (without sum)
of two $6j$'s sharing a common triad
\begin{gather}\label{9jII}
\sum_x (-)^{R+x}(2x+1)
 \begin{Bmatrix}
 a  & b & x \\
 c  & d  & p 
 \end{Bmatrix} 
 \begin{Bmatrix}
 c  & d & x \\
 e  & f  & q
 \end{Bmatrix} 
 \begin{Bmatrix}
 e  & f & x \\
 b & a  & r 
 \end{Bmatrix}\\ \nonumber
\doteq \, 
 \begin{bmatrix}
 a & \, & d & \, & e & \, \\
 \, & p & \,  & q & \, & r \\
 b & \, & c & \, & f & \, \\
 \end{bmatrix}\,=\,
 \begin{Bmatrix}
 p  & q & r \\
 e  & a & d 
 \end{Bmatrix} 
 \begin{Bmatrix}
 p  & q & r \\
 f  & b & c
 \end{Bmatrix}. 
\end{gather}
The content of this formula is however highly non--trivial since it can be 
recognized as the
Biedenharn--Elliott identity relating five $6j$ coefficients.
(Recall that it represents, together with the orthogonality conditions, the
"defining relation" of the hypergeometric polynomial of type $_4F_3$ associated 
with the $6j$, see \cite{Russi} p.295 and \cite{Askey}.)\\ 
The associated Yutsis diagram  is clearly  "separable" on the three lines $p,q,r$
\[
\xy
0*{\bullet}="A";
<-1cm,1.7cm>*{\bullet}="B";
<0cm,3.4cm>*{\bullet}="C";
<2cm,3.4cm>*{\bullet}="D";
<3cm,1.7cm>*{\bullet}="E";
<2cm,0cm>*{\bullet}="F";
"A";"B" **@{-} ?(0.5)*!/_2mm/{e};
"B";"C" **@{-} ?(0.5)*!/_2mm/{d};
"C";"D" **@{-} ?(0.5)*!/_2mm/{p};
"D";"E" **@{-} ?(0.5)*!/_2mm/{c};
"E";"F" **@{-} ?(0.5)*!/_2mm/{f};
"F";"A" **@{-} ?(0.5)*!/_2mm/{r};
"F";"D" **@{-} ?(0.7)*!/_2mm/{b};
"C";"A" **@{-} ?(0.3)*!/_2mm/{a};
"E";"B" **@{-} ?(0.5)*!/_2mm/{q};
<4cm,1.7cm>*{\longrightarrow};
<5cm,1.7cm>*{\bullet}="G";
<6cm,0.7cm>*{\bullet}="L";
<7cm,1.7cm>*{\bullet}="I";
<6cm,2.7cm>*{\bullet}="H";
"G";"H" **@{-} ?(0.5)*!/_2mm/{d};
"G";"I" **@{-} ?(0.7)*!/_2mm/{q};
"H";"I" **@{-} ?(0.5)*!/_2mm/{p};
"I";"L" **@{-} ?(0.5)*!/_2mm/{r};
"L";"G" **@{-} ?(0.5)*!/_2mm/{e};
"L";"H" **@{-} ?(0.3)*!/_2mm/{a};
<8cm,1.7cm>*{\bullet}="M";
<10cm,1.7cm>*{\bullet}="Q";
<9cm,2.7cm>*{\bullet}="N";
<9cm,0.7cm>*{\bullet}="P";
"M";"N" **@{-} ?(0.5)*!/_2mm/{p};
"N";"Q" **@{-} ?(0.5)*!/_2mm/{c};
"Q";"P" **@{-} ?(0.5)*!/_2mm/{f};
"P";"M" **@{-} ?(0.5)*!/_2mm/{r};
"M";"Q" **@{-} ?(0.3)*!/_2mm/{q};
"N";"P" **@{-} ?(0.7)*!/_2mm/{b};
\endxy
\]
where on the right there appear two
Yutsis diagrams of the $6j$ symbol, each  to be thought of as
a complete quadrilateral whose trivalent nodes
correspond to triads of angular momentum variables according
to the convention
\[
\xy
0*{}="A";
<0cm,2cm>*{\bullet}="B";
<2cm,0cm>*{\bullet}="C";
<4cm,2cm>*{\bullet}="D";
<2cm,4cm>*{\bullet}="E";
"B";"E" **@{-} ?(0.5)*!/_2mm/{a};
"E";"D" **@{-} ?(0.5)*!/_2mm/{e};
"D";"C" **@{-} ?(0.5)*!/_2mm/{d};
"C";"B" **@{-} ?(0.5)*!/_2mm/{b};
"E";"C" **@{-} ?(0.7)*!/_3mm/{f};
"B";"D" **@{-} ?(0.3)*!/_2mm/{c};
<6.5cm,2cm>*{\longleftrightarrow\;\;\;\left\{\begin{array}{ccc}
a & b & c\\
d & e & f
\end{array}
\right\}}
\endxy
\]
The two inequivalent ways of applying the Yutsis rule stated
at the beginning of section 1, namely  "inserting
two nodes and joining them", generate in this simplest case two
graphs isomorphic to the $9j$(I)  (left) and  $9j$(II) (right) diagrams
\[
\xy
0*{\bullet}="A";
<3cm,0cm>*{\bullet}="B";
<3cm,3cm>*{\bullet}="C";
<0cm,3cm>*{\bullet}="D";
<0cm,1.5cm>*{\circledcirc}="E";
<3cm,1.5cm>*{\circledcirc}="F";
"B";"A" **@{-} ?(0.5)*!/_2mm/{};
"A";"D" **@{-} ?(0.5)*!/_2mm/{};
"D";"C" **@{-} ?(0.5)*!/_2mm/{};
"C";"B" **@{-} ?(0.5)*!/_2mm/{};
"A";"C" **@{-} ?(0.7)*!/_3mm/{};
"D";"B" **@{-} ?(0.7)*!/_3mm/{};
"E";"F" **@{=} ?(0.7)*!/_3mm/{};
<7cm,0cm>*{\bullet}="A'";
<10cm,0cm>*{\bullet}="B'";
<10cm,3cm>*{\bullet}="C'";
<7cm,3cm>*{\bullet}="D'";
<8.5cm,3cm>*{\circledcirc}="E'";
<10cm,1.5cm>*{\circledcirc}="F'";
"B'";"A'" **@{-} ?(0.5)*!/_2mm/{};
"A'";"D'" **@{-} ?(0.5)*!/_2mm/{};
"D'";"C'" **@{-} ?(0.5)*!/_2mm/{};
"C'";"B'" **@{-} ?(0.5)*!/_2mm/{};
"A'";"C'" **@{-} ?(0.7)*!/_3mm/{};
"D'";"B'" **@{-} ?(0.7)*!/_3mm/{};
"E'";"F'" **@{=} ?(0.7)*!/_3mm/{};
\endxy
\]
The explicit relationship between  such a combinatorial
procedure and the content of the algebraic recursion formulas 
(\ref{9jI}) and (\ref{9jII}) is postposed to section 3, where
the insertion operations will be addressed in the $3nj$(I,II)
(any $n \geq 3$) cases
and related to the general expressions given in
(\ref{RFtypeI}) and (\ref{RFtypeII}).

%%%%%%%%%%%%%%%%%%%%%%%%%%
%%%%%%%%%%%%%%%%%%%%%%%%     SUBSECTION 2.1
%%%%%%%%%%%%%%%%%%%%%%

\subsection{Combinatorial properties of $6j$ and $9j$ graphs}
 
In this section we are going to discuss  properties of
the simplest Yutsis graphs which will be recovered
in an improved way when addressing recursively $3nj$ (I,II) diagrams in
section 3. 
  \begin{itemize}
\item[a)] The diagrams considered so far possess  Hamiltonian
circuits, as shown below.\\ 
\end{itemize}
\[
\xy
0*{}="A";
<0cm,2.8cm>*{}="B";
<2.8cm,2.8cm>*{}="C";
<2.8cm,0cm>*{}="D";
"B";"A" **@{*} ?(0.5)*!/_2mm/{};
"C";"B" **@{*} ?(0.5)*!/_2mm/{};
"D";"C" **@{*} ?(0.5)*!/_2mm/{};
"D";"A" **@{*} ?(0.5)*!/_2mm/{};
"A";"C" **@{-} ?(0.5)*!/_2mm/{};
"D";"B" **@{-} ?(0.5)*!/_2mm/{};
<6cm,0cm>*{}="E";
<7.5cm,0cm>*{}="F";
<8.3cm,1.4cm>*{}="G";
<7.5cm,2.8cm>*{}="H";
<6cm,2.8cm>*{}="I";
<5.2cm,1.4cm>*{}="L";
"E";"F" **@{*} ?(0.5)*!/_2mm/{};
"G";"F" **@{*} ?(0.5)*!/_2mm/{};
"G";"H" **@{*} ?(0.5)*!/_2mm/{};
"H";"I" **@{*} ?(0.5)*!/_2mm/{};
"I";"L" **@{*} ?(0.5)*!/_2mm/{};
"L";"E" **@{*} ?(0.5)*!/_2mm/{};
"L";"G" **@{-} ?(0.5)*!/_2mm/{};
"I";"F" **@{-} ?(0.5)*!/_2mm/{};
"H";"E" **@{-} ?(0.5)*!/_2mm/{};
<11cm,0cm>*{}="E'";
<12.5cm,0cm>*{}="F'";
<13.3cm,1.4cm>*{}="G'";
<12.5cm,2.8cm>*{}="H'";
<11cm,2.8cm>*{}="I'";
<10.2cm,1.4cm>*{}="L'";
"E'";"F'" **@{*} ?(0.5)*!/_2mm/{};
"G'";"F'" **@{*} ?(0.5)*!/_2mm/{};
"G'";"H'" **@{*} ?(0.5)*!/_2mm/{};
"H'";"I'" **@{*} ?(0.5)*!/_2mm/{};
"I'";"L'" **@{*} ?(0.5)*!/_2mm/{};
"L'";"E'" **@{*} ?(0.5)*!/_2mm/{};
"L'";"G'" **@{-} ?(0.5)*!/_2mm/{};
"I'";"E'" **@{-} ?(0.5)*!/_2mm/{};
"H'";"F'" **@{-} ?(0.5)*!/_2mm/{};
\endxy
\]
\vskip 20pt

Recall that a Hamiltonian path in an undirected graph
is a path which visits each vertex exactly once. A Hamiltonian cycle, 
or circuit, is a Hamiltonian path that returns to the starting vertex.
Determining whether such paths and cycles exist in graphs is a typical instance
of  
{\bf NP}--complete problem \cite{GaJo}.\\
As will be shown in section 3, both $3nj$ graphs of type I and  II
admit by construction a Hamiltonian circuit of length
$2n$ for all $n$, bounding a polygonal region 
("plaquette") with the same number of sides
(this is not necessarily  true for other types of
diagrams encountered for $n>4$).
\begin{itemize}
\item[b)] Apart from Hamiltonian circuits, the diagrams
possess other cycles bounding characteristic polygonal
plaquettes. The drawings below display, for each diagram, a typical cycle
of this kind (of course, owing to symmetries of the coefficients,
other cycles of the same shape might have been depicted). 
\end{itemize}
\[
\xy
0*{}="A";
<0cm,2.8cm>*{}="B";
<2.8cm,2.8cm>*{}="C";
<2.8cm,0cm>*{}="D";
"B";"A" **@{-} ?(0.5)*!/_2mm/{};
"C";"B" **@{*} ?(0.5)*!/_2mm/{};
"D";"C" **@{*} ?(0.5)*!/_2mm/{};
"D";"A" **@{-} ?(0.5)*!/_2mm/{};
"A";"C" **@{-} ?(0.5)*!/_2mm/{};
"D";"B" **@{*} ?(0.5)*!/_2mm/{};
<6cm,0cm>*{}="E";
<7.5cm,0cm>*{}="F";
<8.3cm,1.4cm>*{}="G";
<7.5cm,2.8cm>*{}="H";
<6cm,2.8cm>*{}="I";
<5.2cm,1.4cm>*{}="L";
"E";"F" **@{-} ?(0.5)*!/_2mm/{};
"G";"F" **@{*} ?(0.5)*!/_2mm/{};
"G";"H" **@{*} ?(0.5)*!/_2mm/{};
"H";"I" **@{*} ?(0.5)*!/_2mm/{};
"I";"L" **@{-} ?(0.5)*!/_2mm/{};
"L";"E" **@{-} ?(0.5)*!/_2mm/{};
"L";"G" **@{-} ?(0.5)*!/_2mm/{};
"I";"F" **@{*} ?(0.5)*!/_2mm/{};
"H";"E" **@{-} ?(0.5)*!/_2mm/{};
<11cm,0cm>*{}="E'";
<12.5cm,0cm>*{}="F'";
<13.3cm,1.4cm>*{}="G'";
<12.5cm,2.8cm>*{}="H'";
<11cm,2.8cm>*{}="I'";
<10.2cm,1.4cm>*{}="L'";
"E'";"F'" **@{-} ?(0.5)*!/_2mm/{};
"G'";"F'" **@{-} ?(0.5)*!/_2mm/{};
"G'";"H'" **@{*} ?(0.5)*!/_2mm/{};
"H'";"I'" **@{*} ?(0.5)*!/_2mm/{};
"I'";"L'" **@{*} ?(0.5)*!/_2mm/{};
"L'";"E'" **@{-} ?(0.5)*!/_2mm/{};
"L'";"G'" **@{*} ?(0.5)*!/_2mm/{};
"I'";"E'" **@{-} ?(0.5)*!/_2mm/{};
"H'";"F'" **@{-} ?(0.5)*!/_2mm/{};
\endxy
\]
\vskip 20pt 
The number of edges $N^{(n)}_{\text{max}}$ of the "largest", non--Hamiltonian 
circuit in a $3nj$ diagram depends only on $n$ 
(there are only triangular plaquettes  for $n=2$ while there appear quadrilaterals 
for  $n=3$,  pentagons for $n=4$ {\em etc.}, so that $N^{(n)}_{\text{max}}$
$\,=n+1$).
The  $9j$ (II) graph actually contains
also triangular circuits (not highlighted in the above picture) and this feature
is related to the "separability" of this particular
diagram as discussed above. Since the girth
of a graph is defined as the length of the shortest
cycle contained in the graph, the $9j$(II) diagram is
characterized by girth $3$ while the $9j$(I) 
has girth $4$ $=N^{(3)}_{\text{max}}$.\\ 
Generally speaking, the presence
of cycles with a number of edges strictly less than $\,N^{(n)}_{\text{max}}$
is related to the separability of $3nj$ coefficients into a
product (no sum) of lower order coefficients, namely to  "trivial" 
configurations.
\begin{itemize}
\item[c)] The existence of circuits (Hamiltonian
or not) can be assumed as a "guiding principle" in the search for
asymptotic expansions of the $3nj$ coefficients 
by letting the edges belonging to the circuit to become "large" while
keeping "quantum" the others. This is  what
has been already done for the $6j$ in 
\cite{AqHaLi1},
for the $9j$ in \cite{AnAqFe} and general
results on the so--called phenomenon of  "disentangling" of networks 
will be discussed in section 4 below. \\
\item[d)] By resorting to  notions from topological graph theory
(see section 3) it is possible to
classify Yutsis graphs associated with $3nj$(I, II) diagrams
according to the possibility of  "embedding" them on closed
surfaces. 
\end{itemize}
Consider in particular the graphs of the $6j$ (to be 
classified as type II!) and of the $9j$ (II) (the separable one).
They are both embeddable on a 2--sphere since they represent  
(the surfaces of) a tetrahedron
and a triangular prism, respectively
\[
\xy
0*{}="A";
<0cm,1cm>*{\bullet}="B";
<1cm,3cm>*{\bullet}="C";
<3cm,1cm>*{\bullet}="D";
<1cm,0cm>*{\bullet}="E";
"B";"C" **@{-} ?(0.5)*!/_2mm/{};
"C";"D" **@{-} ?(0.5)*!/_2mm/{};
"D";"E" **@{-} ?(0.5)*!/_2mm/{};
"B";"D" **@{.} ?(0.5)*!/_2mm/{};
"E";"B" **@{-} ?(0.5)*!/_2mm/{};
"C";"E" **@{-} ?(0.5)*!/_2mm/{};
<5cm,0cm>*{\bullet}="F";
<7cm,3.4cm>*{\bullet}="G";
<9cm,0cm>*{\bullet}="H";
<6cm,0.6cm>*{\bullet}="I";
<8cm,0.6cm>*{\bullet}="L";
<7cm,2.2cm>*{\bullet}="M";
"F";"G" **@{-} ?(0.5)*!/_2mm/{};
"G";"H" **@{-} ?(0.5)*!/_2mm/{};
"H";"F" **@{-} ?(0.5)*!/_2mm/{};
"I";"M" **@{-} ?(0.5)*!/_2mm/{};
"M";"L" **@{-} ?(0.5)*!/_2mm/{};
"I";"L" **@{-} ?(0.5)*!/_2mm/{};
"F";"I" **@{-} ?(0.5)*!/_2mm/{};
"L";"H" **@{-} ?(0.5)*!/_2mm/{};
"G";"M" **@{-} ?(0.5)*!/_2mm/{};
\endxy
\]
On the other hand, the $9j$(I) diagram is a genus--1
or "toroidal" graph, as can be inferred by noting that
this graph is isomorphic to $K_{3,3}$ (the bipartite graph
on 6 nodes, or "utility graph"). To make this point explicit,
recall that the torus $T^2$ can be thought of as a
rectangle in the plane with   opposite edges pairwise identified
without twisting. The existence of a Hamiltonian circuit
connecting all of the six nodes  
is crucial here: the circuit with its nodes is drawn as an 
horizontal  straight line 
inside the rectangle and represents a longitudinal  path on the torus 
(which is topologically a circle owing to identification 
of  opposite points on  vertical boundaries). 
The other connections
are established in such a way that the connectivity of the original
diagram is recovered (points to be identified pairwise have the same
horizontal coordinates on the upper and lower boundaries).
\[
\xy
0*{}="A";
<3cm,0cm>*{\diamond}="B";
<4cm,0cm>*{\diamond}="C";
<5cm,0cm>*{\diamond}="D";
<8cm,0cm>*{}="E";
<8cm,2cm>*{\circ}="F";
<7cm,2cm>*{\bullet}="G";
<6cm,2cm>*{\bullet}="H";
<5cm,2cm>*{\bullet}="I";
<3cm,2cm>*{\bullet}="L";
<2cm,2cm>*{\bullet}="M";
<1cm,2cm>*{\bullet}="N";
<0cm,2cm>*{\circ}="O";
<0cm,4cm>*{}="P";
<3cm,4cm>*{\diamond}="Q";
<4cm,4cm>*{\diamond}="R";
<5cm,4cm>*{\diamond}="S";
<8cm,4cm>*{}="T";
"A";"E" **@{.} ?(0.5)*!/_2mm/{};
"F";"O" **@{-} ?(0.5)*!/_2mm/{};
"P";"T" **@{.} ?(0.5)*!/_2mm/{};
"A";"P" **@{--} ?(0.5)*!/_2mm/{};
"E";"T" **@{--} ?(0.5)*!/_2mm/{};
"B";"I" **@{-} ?(0.5)*!/_2mm/{};
"C";"H" **@{-} ?(0.5)*!/_2mm/{};
"D";"G" **@{-} ?(0.5)*!/_2mm/{};
"N";"Q" **@{-} ?(0.5)*!/_2mm/{};
"M";"R" **@{-} ?(0.5)*!/_2mm/{};
"L";"S" **@{-} ?(0.5)*!/_2mm/{};
\endxy
\]

In section 3.2 the embedding of both $9j$(I)  
and any other $3nj$(I) diagram into the projective 
space will be addressed.

%%%%%%%%%%%%%%%%%%%%%%%%%%%%%%%%%%%%%
%%%%%%%%%%%%%%%%%%%%%%%%%%            SUBSECTION 2.2
%%%%%%%%%%%%%%%%%%%%%%%%%%%%%%%%%%%

\subsection{Recursion relation for the $9j$ symbol}

According to remark iii) of section 1, the starting point is 
the generating formula for a type I $3nj$ coefficient given in (\ref{RFtypeI}).
It can be used to get an expression involving (sums of) products
of two $3(n-1)j$ and four $6j$ as in (\ref{3njsymm}).
 The explicit form of the latter in the present case reads
 (\cite{Russi} p. 345; the same in \cite{YuLeVa} p.175)
 \begin{gather}\label{12jsymm}
(-)^{a+f+b+j} \,\sum_x \,(-)^{2x} (2x+1)
\begin{Bmatrix}
 a & b & x \\
 d & e & f \\
 g & h & j 
 \end{Bmatrix}
 \begin{Bmatrix}
 a & b & x \\
 c & \lambda & a' 
 \end{Bmatrix}\;
 \begin{Bmatrix}
 j & f & x \\
 \lambda & c & f' 
 \end{Bmatrix} \\ \nonumber 
=\,(-)^{a'+f'+g+e} \,\sum_y\,(-)^{2y} (2y+1)
\begin{Bmatrix}
 a' & b & c \\
 y & e & f' \\
 g & h & j 
 \end{Bmatrix}
 \begin{Bmatrix}
 f' & e & y \\
 d & \lambda & f 
 \end{Bmatrix}\;
 \begin{Bmatrix}
 g & a' & y \\
 \lambda & d & a 
 \end{Bmatrix}
\end{gather}
Specifying $\lambda =1$, $a'=a$ and $f'=f$ and using explicit expressions for the
$6j$ symbols, the above formula gives a
5--term recursion relation for the $9j$ 
(\cite{Russi}, (10.5.3) p. 347; the same in \cite{YuBa} p.177), 
namely
%%RR 9J REWRITTEN
 \begin{gather}\label{RR9j1}
 \frac{\mathcal{A}_{c+1}(ab,fj)}
{(c+1)(2c+1)}
\left\{\begin{smallmatrix}
 c+1 & a & b \\
 f & d & e \\
 j & g & h 
 \end{smallmatrix} \right\} +
 \frac{\mathcal{A}_{c}(ab,fj)}
{c(2c+1)}
\left\{\begin{smallmatrix}
 c-1 & a & b \\
 f & d & e \\
 j & g & h 
 \end{smallmatrix}\right\}
+\mathcal{C}(a,b,c,f,j)
\left\{\begin{smallmatrix}
 c & a & b \\
 f & d & e \\
 j & g & h 
 \end{smallmatrix}\right\}
\\ \nonumber
 =\frac{\mathcal{A}_{d+1}(ef,ag)}
{(d+1)(2d+1)}
\left\{\begin{smallmatrix}
  c & a & b \\
 f & d+1 & e \\
 j & g & h 
 \end{smallmatrix}\right\}
 +
 \frac{\mathcal{A}_{d}(ef,ag)}
{d(2d+1)}
\left\{\begin{smallmatrix}
  c & a & b \\
 f & d-1 & e \\
 j & g & h 
 \end{smallmatrix}\right\} +
  \mathcal{D}(a,d,g,e,f)
\left\{\begin{smallmatrix}
 c & a & b \\
 f & d & e \\
 j & g & h 
 \end{smallmatrix}\right\},
\end{gather}
where symmetries of the $9j$ have been used.
The functions in front of the $9j$ coefficients
are given explicitly by
\begin{gather}\label{RRA}
\mathcal{A}_{q}(pr,st)=
[(-p+r+q)(p-r+q)(p+r-q+1)(p+r+q+1)\\ \nonumber
\times (-s+t+q)(s-t+q)(s+t-q+1)(s+t+q+1)
]^{\tfrac{1}{2}}
\end{gather}

\begin{equation}\label{RRC}
\mathcal{C}(a,b,c,f,j)=
\frac{[a(a+1)- b(b+1)+c(c+1)]
[c(c+1)+f(f+1)-j(j+1)]}
{c(c+1)}
\end{equation}

\begin{equation}\label{RRB}
\mathcal{D}(a,d,g,e,f)=
\frac{[a(a+1)- d(d+1)-g(g+1)]
[d(d+1)- e(e+1)+f(f+1)]}
{d(d+1)}.
\end{equation}

The meaning of these quantities is geometrical, as happens in the
3--term recurrence relation for the $6j$ symbol \cite{PoRe}, \cite{BiLo9}
(topic 9), \cite{Russi}. 
(Note however that the simplest derivation can be found in \cite{Nev},
where it was shown that only $6j$'s where one of the entries is 1
are involved, as happens also here for the $9j$.)
In particular,
$\mathcal{A}_{q}(pr,st)$ represents the squared area of the
quadrilateral bounded by $p,r,s,t$ and made of two triangles
meeting at $q$, while $\mathcal{C}$ and
$\mathcal{D}$ are cosines of the dihedral angles at
$c$ and $d$, respectively (see \cite{Ragni}, section III
 for more details).
Each line in (\ref{RR9j1}) closely resembles
the structure of the recurrence formula of the $6j$,
the main difference being that here there are
two running variables, $(c,d)$ instead of one
(the other spins play the role of parameters).
Notice that the entry labeled by $h$
is not active, but we might   write down
a similar expression involving either the pair $(c,h)$
or $(d,h)$, and finally collect all of them
in a 7--term relation in the three running
variables $c,d,h$, as we are going to  sketch below. 

By introducing the shorthand notation
\begin{equation*}
\left\{\begin{smallmatrix}
 c & a & b \\
 f & d & e \\
 j & g & h 
 \end{smallmatrix}\right\}\,\doteq \,
 \mathcal{N}(c,d,h);\;\;\;\;\mathcal{C}(a,b,c,f,j)\,\doteq \,
 \tilde{\mathcal{C}}_c
\end{equation*}
\begin{equation}\label{notation}
\mathcal{D}(a,d,g,e,f) \,\doteq \,
 \tilde{\mathcal{D}}_d;
\;\;\;\;\mathcal{H}(b,e,h,g,j)\,\doteq \,
 \tilde{\mathcal{H}}_h,
\end{equation}
where $\tilde{\mathcal{H}}_h$ represents
the  function to be associated with  the third running variable,
the complete recursion equation can be casted  in the form
\begin{equation*}
\tilde{\mathcal{A}}_{c+1} \, \mathcal{N}(c+1,d,h) +
\tilde{\mathcal{A}}_{c} \, \mathcal{N}(c-1,d,h) +
\tilde{\mathcal{C}}_c \, \mathcal{N}(c,d,h)
\end{equation*}
\begin{equation}\label{RR9j2}
=\,\tilde{\mathcal{A}}_{d+1}\, \mathcal{N}(c,d+1,h) +
\tilde{\mathcal{A}}_{d} \, \mathcal{N}(c,d-1,h) +
\tilde{\mathcal{D}}_c \, \mathcal{N}(c,d,h)
\end{equation}
\begin{equation*}
=\,\tilde{\mathcal{A}}_{h+1} \, \mathcal{N}(c,d,h+1) +
\tilde{\mathcal{A}}_{h} \, \mathcal{N}(c,d,h-1) +
\tilde{\mathcal{H}}_h \, \mathcal{N}(c,d,h)
\end{equation*}
(the symbols $\tilde{\mathcal{A}}$ include the proper normalization factors
defined as in (\ref{RR9j1})).\\
In the (triple) asymptotic limit $(c,d,h)  \gg 1 \rightarrow (x,y,z)$
the  "difference" equation above becomes a partial
differential equation
for the function $\mathcal{N}(x,y,z)$ 
\begin{equation*}
\left[\Delta_x^2 -2 \cos\theta_x +2\right]
\mathcal{N}(x,y,z) = 
 \left[\Delta_y^2 -2 \cos\theta_y +2\right]\mathcal{N}(x,y,z)
\end{equation*} 
 \begin{equation}\label{PDE9j}
= \left[\Delta_z^2 -2 \cos\theta_z +2\right]
\mathcal{N}(x,y,z)
\end{equation}
where $\Delta_x^2$, $\Delta_y^2$,  $\Delta_z^2$ 
denote second (partial) derivatives with respect to the 
continuous variables $x,y,z$ while $\theta_x$,
$\theta_y$, $\theta_z$
are angles whose cosines are defined in terms of both $x,y,z$
and the other parameters of the original $9j$ coefficient.
The derivation of each member of (\ref{PDE9j})
can be carried out as in \cite{ScGo} and we refer
the reader to \cite{Ragni} for details (see in particular
their Eq. 22). However, in the present case the proper
domain of the function $\mathcal{N}(x,y,z)$ 
is the Euclidean space $\mathbb{R}^3$ 
and $(\Delta_x^2,\Delta_y^2,\Delta_z^2 ) \equiv \Delta^2$  is the 3--dimensional
Laplace operator.
Moreover, if we equate  each member of (\ref{PDE9j}) separately to the same function 
$\mathcal{N}$ times a
constant (to be determined), we would get  a set of coupled eigenvalue equations, and the
allowed values for the separation constant would generate the eigenvalue
spectrum. Two more remarks are in order
\begin{itemize}
\item[(i)] By setting   $h=0$ in (\ref{RR9j1}) the conditions
$b=e$ and $j=g$ follow necessarily. 
Then, according to remark iii) and iv) of section 1,
we would  see how the 5--term recursion for the $9j$
simplifies for the $6j$ into a 5-term
recursion and thus, as we have just seen,
in a two--variable difference
equation which, in the asymptotic limit, 
leads to a two--dimensional coupled partial
differential problem.
Though such kind of procedure seems rather trivial because we already know 
that the $6j$ satisfied a 3--term recurrence  for which  the "separation constant"
is given, this point illustrates the "dual role" of, say,  
$c$ and $d$, and the interest
of studying the $6j$ symbol as a function of two variables at 
fixed values of the other four entries
(see  \cite{AqHaLi1}).
\item[(ii)] The above procedure, when applied to higher $3nj$ 
symbols, can be shown to lead to
exactly $n$ coupled equations  in $n$ variables for symbols of 
the first and second kind, and for
highly symmetric ones, such as the $15j$ of the fifth type,
whose Yutsis representation is the Petersen graph. Less symmetric 
cases might be dealt with according to the degree
of their edge--transitivity.
\end{itemize}

More generally, such  asymptotic  techniques based on recurrence formulas
--developed here for the $9j$-- represent the starting point for 
further developments concerning Rodriguez--type formulas (the defining
differential relations of families of orthogonal polynomials, see \cite{Askey}
\cite{NiSuUv} for 1--variable  polynomials),
recursion relations  \cite{BrCaDr}
and their relationships with
the issue of separability of Schr\"odinger equation in many--body
quantum systems, see \cite{Ragni} and references therein.
Apart from a short account in  \cite{NiSuUv} 
(Ch. 5, Sec. 4), such kind of analysis
in the case of 2--variable orthogonal polynomials of 
hypergeometric type --as the $9j$ happens to be--
seems to have been addressed only quite recently in the mathematical literature
\cite{ArCoVa}.
We argue that results we are going to review in section 4.1
on particular asymptotic expansions of the $9j$ and 
the novel general scheme for asymptotic disentangling of
$3nj$ discussed in section 4.2
might contribute to shed light also on the generation
of hierarchies of multi--variable special functions of the hypergeometric type. 
Work is in progress in this direction.

%%%%%%%%%%%%%%%%%%%%%%%%%%
%%%%%%%%%%%%%%%%%%%%%           SECTION 3: INTRODUCTORY PART
%%%%%%%%%%%%%%%%%%%%%%%%%%
%%%%%%%%%%%%%%%%%%%%%%%%%%

\section{Genesis of $3nj$ diagrams}

Looking at Yutsis diagrams as particular families of (unoriented) graphs,
namely collections of nodes (vertices) and links (edges)
connecting pairs of nodes, 
it is worth to introduce a few notions and definitions
from topological graph theory \cite{GrTu}.

A graph is embeddable on a surface $\Sigma$
if its vertices and edges can be arranged on it without any crossing.
The "genus" of a graph is the {\em lowest} genus of any surface on which the
graph is embedded.
Recall that closed and {\em orientable} surfaces $\Sigma$
are topologically classified  by their Euler number
$\chi (\Sigma)$ $=2-2g$, where the genus $g$ is the number
of "handles" or "holes"; 
then $\chi=2$ ($g=0$, namely no handles) gives the 2--sphere $S^2$,
$\chi=0$ ($g=1$) gives the 2--torus $T^2$ (whose presentation has
been  shown at the end of section 2.1), 
while the cases $g=3,4\dots$ correspond to double, triple, ... torus  $\Sigma_g$,
all characterized by an even negative $\chi$.
A planar graph is one which can be drawn on the Euclidean plane or,
{\em equivalently}, on the sphere $S^2$ without any crossing, namely
it is has genus $0$ or $\chi =2$.
\vskip 5pt
The issue of the classification of Yutsis graphs 
is a longstanding problem \cite{BiLo9,Belgi}  and
in particular 
such a "topological" viewpoint has been  addressed in \cite{KrLo}.
In this section we are going to complete  the characterization
of all $3nj$(I,II) diagrams by resorting to the analysis of
the combinatorial and topological  contents of
the "insertion operations" in connection with the
recurrence formulas (\ref{RFtypeI}) (type I) and (\ref{RFtypeII})
(type II). Moreover, the existence of Hamiltonian cycles (defined below)
in any of such diagrams will be the starting point of the
analysis of asymptotic disentangling developed in section 4.
Finally, a few remarks on the application of our recursive
procedure to other types of $3nj$ diagrams are briefly discussed. 
 
We start from the analysis of type--II graphs since 
they share  simpler features.

%%%%%%%%%%%%%%%%%%%%%%%%
%%%%%%%%%%%%%%%%%%%%%%       SECTION 3.1 (PROP.1 and 2)
%%%%%%%%%%%%%%%%%%%%%%%%%%%

\subsection{$3nj$(II) diagrams}

{\bf Proposition 1}.\\  
{\em Any $3nj$ diagram of type II is planar (or genus--0).}

\vskip 12pt

\noindent Recall from section 2.1 that both the $6j$ (tetrahedron) and the
$9j$(II) (triangular prism) graphs are easily recognized as planar. 
Comparing the original diagrams up to the $18j$
given in \cite{YuLeVa,YuBa} it turns out that
a convenient and unified way of drawing all $3nj$(II)
is through Schlegel diagrams of prisms based on polygons with 
an increasing number of sides. Below the 
square ($12j$), pentagonal ($15j$), hexagonal ($18j$)
prisms of this sort are depicted, while the triangular prism associated
with the $9j$(I) has been already shown at the end of section 2.1.
\[
\xy
0*{\bullet}="A";
<3cm,0cm>*{\bullet}="B";
<3cm,3cm>*{\bullet}="C";
<0cm,3cm>*{\bullet}="D";
<1cm,1cm>*{\bullet}="A'";
<2cm,1cm>*{\bullet}="B'";
<2cm,2cm>*{\bullet}="C'";
<1cm,2cm>*{\bullet}="D'";
"A";"B" **@{-} ?(0.5)*!/_2mm/{};
"B";"C" **@{-} ?(0.5)*!/_2mm/{};
"C";"D" **@{-} ?(0.5)*!/_2mm/{};
"D";"A" **@{-} ?(0.5)*!/_2mm/{};
"A";"A'" **@{-} ?(0.5)*!/_2mm/{};
"B";"B'" **@{-} ?(0.5)*!/_2mm/{};
"C";"C'" **@{-} ?(0.5)*!/_2mm/{};
"D";"D'" **@{-} ?(0.5)*!/_2mm/{};
"A'";"B'" **@{-} ?(0.5)*!/_2mm/{};
"B'";"C'" **@{-} ?(0.5)*!/_2mm/{};
"C'";"D'" **@{-} ?(0.5)*!/_2mm/{};
"D'";"A'" **@{-} ?(0.5)*!/_2mm/{};
<5.7cm,0cm>*{\bullet}="E";
<8.2cm,0cm>*{\bullet}="F";
<9cm,2.3cm>*{\bullet}="G";
<7cm,3.8cm>*{\bullet}="H";
<5cm,2.3cm>*{\bullet}="I";
<6.3cm,0.8cm>*{\bullet}="E'";
<7.5cm,0.8cm>*{\bullet}="F'";
<7.9cm,2cm>*{\bullet}="G'";
<7cm,2.7cm>*{\bullet}="H'";
<6cm,2cm>*{\bullet}="I'";
"E";"F" **@{-} ?(0.5)*!/_2mm/{};
"F";"G" **@{-} ?(0.5)*!/_2mm/{};
"G";"H" **@{-} ?(0.5)*!/_2mm/{};
"H";"I" **@{-} ?(0.5)*!/_2mm/{};
"I";"E" **@{-} ?(0.5)*!/_2mm/{};
"E'";"F'" **@{-} ?(0.5)*!/_2mm/{};
"F'";"G'" **@{-} ?(0.5)*!/_2mm/{};
"G'";"H'" **@{-} ?(0.5)*!/_2mm/{};
"H'";"I'" **@{-} ?(0.5)*!/_2mm/{};
"I'";"E'" **@{-} ?(0.5)*!/_2mm/{};
"E";"E'" **@{-} ?(0.5)*!/_2mm/{};
"F";"F'" **@{-} ?(0.5)*!/_2mm/{};
"G";"G'" **@{-} ?(0.5)*!/_2mm/{};
"H";"H'" **@{-} ?(0.5)*!/_2mm/{};
"I";"I'" **@{-} ?(0.5)*!/_2mm/{};
<12cm,0cm>*{\bullet}="L";
<14cm,0cm>*{\bullet}="M";
<15cm,1.6cm>*{\bullet}="N";
<14cm,3.4cm>*{\bullet}="O";
<12cm,3.4cm>*{\bullet}="P";
<11cm,1.6cm>*{\bullet}="Q";
<12.5cm,0.8cm>*{\bullet}="L'";
<13.4cm,0.8cm>*{\bullet}="M'";
<14cm,1.6cm>*{\bullet}="N'";
<13.4cm,2.5cm>*{\bullet}="O'";
<12.5cm,2.5cm>*{\bullet}="P'";
<12cm,1.6cm>*{\bullet}="Q'";
"L";"M" **@{-} ?(0.5)*!/_2mm/{};
"M";"N" **@{-} ?(0.5)*!/_2mm/{};
"N";"O" **@{-} ?(0.5)*!/_2mm/{};
"O";"P" **@{-} ?(0.5)*!/_2mm/{};
"P";"Q" **@{-} ?(0.5)*!/_2mm/{};
"Q";"L" **@{-} ?(0.5)*!/_2mm/{};
"L'";"M'" **@{-} ?(0.5)*!/_2mm/{};
"M'";"N'" **@{-} ?(0.5)*!/_2mm/{};
"N'";"O'" **@{-} ?(0.5)*!/_2mm/{};
"O'";"P'" **@{-} ?(0.5)*!/_2mm/{};
"P'";"Q'" **@{-} ?(0.5)*!/_2mm/{};
"Q'";"L'" **@{-} ?(0.5)*!/_2mm/{};
"L";"L'" **@{-} ?(0.5)*!/_2mm/{};
"M";"M'" **@{-} ?(0.5)*!/_2mm/{};
"N";"N'" **@{-} ?(0.5)*!/_2mm/{};
"O";"O'" **@{-} ?(0.5)*!/_2mm/{};
"P";"P'" **@{-} ?(0.5)*!/_2mm/{};
"Q";"Q'" **@{-} ?(0.5)*!/_2mm/{};
\endxy
\]
Every $3nj$(II) has a Hamiltonian circuit 
of length $2n$, as can be easily inferred from the two
samples shown below. 
\[
\xy
0*{}="A";
<3cm,0cm>*{}="B";
<3cm,3cm>*{}="C";
<0cm,3cm>*{}="D";
<1cm,1cm>*{}="A'";
<2cm,1cm>*{}="B'";
<2cm,2cm>*{}="C'";
<1cm,2cm>*{}="D'";
"A";"B" **@{-} ?(0.5)*!/_2mm/{};
"B";"C" **@{*} ?(0.5)*!/_2mm/{};
"C";"D" **@{*} ?(0.5)*!/_2mm/{};
"D";"A" **@{*} ?(0.5)*!/_2mm/{};
"A";"A'" **@{*} ?(0.5)*!/_2mm/{};
"B";"B'" **@{*} ?(0.5)*!/_2mm/{};
"C";"C'" **@{-} ?(0.5)*!/_2mm/{};
"D";"D'" **@{-} ?(0.5)*!/_2mm/{};
"A'";"B'" **@{-} ?(0.5)*!/_2mm/{};
"B'";"C'" **@{*} ?(0.5)*!/_2mm/{};
"C'";"D'" **@{*} ?(0.5)*!/_2mm/{};
"D'";"A'" **@{*} ?(0.5)*!/_2mm/{};
<5.7cm,0cm>*{}="E";
<8.2cm,0cm>*{}="F";
<9cm,2.3cm>*{}="G";
<7cm,3.8cm>*{}="H";
<5cm,2.3cm>*{}="I";
<6.3cm,0.8cm>*{}="E'";
<7.5cm,0.8cm>*{}="F'";
<7.9cm,2cm>*{}="G'";
<7cm,2.7cm>*{}="H'";
<6cm,2cm>*{}="I'";
"E";"F" **@{-} ?(0.5)*!/_2mm/{};
"F";"G" **@{*} ?(0.5)*!/_2mm/{};
"G";"H" **@{*} ?(0.5)*!/_2mm/{};
"H";"I" **@{*} ?(0.5)*!/_2mm/{};
"I";"E" **@{*} ?(0.5)*!/_2mm/{};
"E'";"F'" **@{-} ?(0.5)*!/_2mm/{};
"F'";"G'" **@{*} ?(0.5)*!/_2mm/{};
"G'";"H'" **@{*} ?(0.5)*!/_2mm/{};
"H'";"I'" **@{*} ?(0.5)*!/_2mm/{};
"I'";"E'" **@{*} ?(0.5)*!/_2mm/{};
"E";"E'" **@{*} ?(0.5)*!/_2mm/{};
"F";"F'" **@{*} ?(0.5)*!/_2mm/{};
"G";"G'" **@{-} ?(0.5)*!/_2mm/{};
"H";"H'" **@{-} ?(0.5)*!/_2mm/{};
"I";"I'" **@{-} ?(0.5)*!/_2mm/{};
\endxy
\]
Other non--Hamiltonian cycles
of increasing length appear as well, and it turns out
that the girth number
--the length of the shortest
cycle-- 
is exactly $4$ for all of these graphs
(recall that the property of having girth $3$, namely
the existence of triangular circuit(s) implies separability
of the diagram into lower order diagrams and as such has been
already ruled out, {\em cfr.} the recursion  rule
stated at the beginning of section 1).

\vskip 12pt

\noindent {\bf Proposition 2}.\\
{\em There exists (up to symmetry) one kind
of insertion operation, denoted by
 $\mathfrak{I}_{\square}\,$, that generates
the $3nj$(II) from the $3(n-1)j$(II) for
any $n\geq 3$ according to the recurrence formula
(\ref{RFtypeII}). Moreover $\mathfrak{I}_{\square}$
is topology--preserving, namely all  graphs 
derived in such a way stay planar.}  

\vskip 12pt

\noindent The proof can be carried out by resorting inductively to Yutsis
graphical method which mimics step--by--step
algebraic manipulations to be performed on equations. The figures below
deal with the generation of the $12j$(II)
from the $9j$(II) and make it manifest the effective equivalence
between the combinatorial recursion method and the algebraic recurrence formula.\\ 
In the first figure the
recursion rule "add two vertices and join them"
(pictorially  $\circledcirc\negthinspace=\negthinspace\circledcirc$) 
is applied on any one of the lateral faces of the triangular
prism.\\
 In the second one the
content of the algebraic recurrence formula (\ref{RFtypeII}) 
for $3nj$(II), $n=3$, is displayed: the summation
over $x$ implies  the cancellation of both edges labeled by $x$
and nodes which lose an incident edge; in the last step
edges with the same labels are joined again to give the final closed
diagram.\\
Note that the two diagrams of the $9j$(II)
on the left--hand side of each picture
are isomorphic, as well as the
two resulting graphs  of the $12j$(II)
on the right--hand sides. 
\[
\xy
0*{\bullet}="A";
<2cm,-2cm>*{\bullet}="B";
<2cm,2cm>*{\bullet}="C";
<0.4cm,0cm>*{\bullet}="A'";
<1.3cm,-0.5cm>*{\bullet}="B'";
<1.3cm,0.5cm>*{\bullet}="C'";
"A";"B" **@{-} ?(0.5)*!/_2mm/{};
"B";"C" **@{-} ?(0.5)*!/_2mm/{};
"C";"A" **@{-} ?(0.5)*!/_2mm/{};
"A";"A'" **@{-} ?(0.5)*!/_2mm/{};
"B";"B'" **@{-} ?(0.5)*!/_2mm/{};
"C";"C'" **@{-} ?(0.5)*!/_2mm/{};
"A'";"B'" **@{-} ?(0.5)*!/_2mm/{};
"B'";"C'" **@{-} ?(0.5)*!/_2mm/{};
"C'";"A'" **@{-} ?(0.5)*!/_2mm/{};
"B'";"A'" **@{} ?(0.5)*!/_2mm/{};
<3cm,0cm>*{\rightarrow};
<4cm,0cm>*{\bullet}="D";
<6cm,-2cm>*{\bullet}="E";
<6.5cm,0cm>*{\circledcirc}="F";
<6cm,2cm>*{\bullet}="G";
<4.4cm,0cm>*{\bullet}="D'";
<5cm,-0.5cm>*{\bullet}="E'";
<5.5cm,0cm>*{\circledcirc}="F'";
<5cm,0.5cm>*{\bullet}="G'";
"D";"G" **@{-} ?(0.5)*!/_2mm/{};
"G";"F" **@{-} ?(0.5)*!/_2mm/{};
"F";"E" **@{-} ?(0.5)*!/_2mm/{};
"E";"D" **@{-} ?(0.5)*!/_2mm/{};
"D";"D'" **@{-} ?(0.5)*!/_2mm/{};
"G";"G'" **@{-} ?(0.5)*!/_2mm/{};
"F";"F'" **@{=} ?(0.5)*!/_2mm/{};
"E";"E'" **@{-} ?(0.5)*!/_2mm/{};
"G'";"D'" **@{-} ?(0.5)*!/_2mm/{};
"G'";"F'" **@{-} ?(0.5)*!/_2mm/{};
"F'";"E'" **@{-} ?(0.5)*!/_2mm/{};
"E'";"D'" **@{-} ?(0.5)*!/_2mm/{};
"G";"E" **@{.} ?(0.5)*!/_2mm/{};
"G'";"E'" **@{.} ?(0.5)*!/_2mm/{};
<8cm,0cm>*{\rightarrow};
<11cm,-2cm>*{\bullet}="H";
<13cm,0cm>*{\bullet}="I";
<11cm,2cm>*{\bullet}="L";
<9cm,0cm>*{\bullet}="M";
<11cm,-1cm>*{\bullet}="H'";
<12cm,0cm>*{\bullet}="I'";
<11cm,1cm>*{\bullet}="L'";
<10cm,0cm>*{\bullet}="M'";
"M";"L" **@{-} ?(0.5)*!/_2mm/{};
"I";"L" **@{-} ?(0.5)*!/_2mm/{};
"I";"H" **@{-} ?(0.5)*!/_2mm/{};
"M";"H" **@{-} ?(0.5)*!/_2mm/{};
"M'";"L'" **@{-} ?(0.5)*!/_2mm/{};
"I'";"L'" **@{-} ?(0.5)*!/_2mm/{};
"I'";"H'" **@{-} ?(0.5)*!/_2mm/{};
"M'";"H'" **@{-} ?(0.5)*!/_2mm/{};
"M";"M'" **@{-} ?(0.5)*!/_2mm/{};
"L";"L'" **@{-} ?(0.5)*!/_2mm/{};
"I";"I'" **@{-} ?(0.5)*!/_2mm/{};
"H";"H'" **@{-} ?(0.5)*!/_2mm/{};
<5.5cm,-3cm>*{\text{{\bf Combinatorial recursion rule} (type II)}};
\endxy
\]

\vskip 12pt
%%%%%%%%%%%%%%%%%%%%%%%%
\[
\xy
0*{}="A";
<1cm,0cm>*{\bullet}="B";
<2cm,1cm>*{\bullet}="C";
<3cm,1cm>*{\bullet}="D";
<4cm,0cm>*{\bullet}="E";
<3cm,-1cm>*{\bullet}="F";
<2cm,-1cm>*{\bullet}="G";
"B";"C" **@{-} ?(0.5)*!/_2mm/{};
"C";"D" **@{-} ?(0.5)*!/_2mm/{};
"D";"E" **@{-} ?(0.5)*!/_2mm/{};
"E";"F" **@{-} ?(0.5)*!/_2mm/{};
"G";"F" **@{-} ?(0.5)*!/_2mm/{x};
"B";"G" **@{-} ?(0.5)*!/_2mm/{};
"B";"E" **@{-} ?(0.5)*!/_2mm/{};
"C";"G" **@{-} ?(0.5)*!/_2mm/{};
"D";"F" **@{-} ?(0.5)*!/_2mm/{};
<0cm,-1.5cm>*{\sum_x};
<4.5cm,-1.5cm>*{=};
<1cm,-2.5cm>*{\bullet}="B'";
<1.5cm,-1.5cm>*{\bullet}="C'";
<2cm,-2.5cm>*{\bullet}="D'";
<1.5cm,-3.5cm>*{\bullet}="E'";
"B'";"C'" **@{-} ?(0.5)*!/_2mm/{};
"C'";"D'" **@{-} ?(0.5)*!/_2mm/{x};
"D'";"E'" **@{-} ?(0.5)*!/_2mm/{};
"E'";"B'" **@{-} ?(0.5)*!/_2mm/{};
"C'";"E'" **@{-} ?(0.5)*!/_2mm/{};
"B'";"D'" **@{-} ?(0.5)*!/_2mm/{};
<3cm,-2.5cm>*{\bullet}="B''";
<3.5cm,-1.5cm>*{\bullet}="C''";
<4cm,-2.5cm>*{\bullet}="D''";
<3.5cm,-3.5cm>*{\bullet}="E''";
"B''";"C''" **@{-} ?(0.5)*!/_2mm/{x};
"C''";"D''" **@{-} ?(0.5)*!/_2mm/{};
"D''";"E''" **@{-} ?(0.5)*!/_2mm/{};
"E''";"B''" **@{-} ?(0.5)*!/_2mm/{};
"C''";"E''" **@{-} ?(0.5)*!/_2mm/{};
"B''";"D''" **@{-} ?(0.5)*!/_2mm/{};
<6cm,0cm>*{\bullet}="H";
<7cm,1cm>*{\bullet}="I";
<8cm,1cm>*{\bullet}="L";
<9cm,0cm>*{\bullet}="K";
<6cm,-1cm>*{}="H'";
<7cm,-1cm>*{}="I'";
<8cm,-1cm>*{}="L'";
<9cm,-1cm>*{}="K'";
"H'";"H" **@{-} ?(0.5)*!/_2mm/{a};
"I'";"I" **@{-} ?(0.3)*!/_2mm/{a'};
"L";"L'" **@{-} ?(0.7)*!/_2mm/{b};
"K";"K'" **@{-} ?(0.5)*!/_2mm/{b'};
"H";"I" **@{-} ?(0.5)*!/_2mm/{};
"L";"I" **@{-} ?(0.5)*!/_2mm/{};
"K";"L" **@{-} ?(0.5)*!/_2mm/{};
"H";"K" **@{-} ?(0.5)*!/_2mm/{};
<5.5cm,-1.5cm>*{}="H''";
<6cm,-1.5cm>*{}="I''";
<9cm,-1.5cm>*{}="L''";
<9.5cm,-1.5cm>*{}="K''";
<5.5cm,-2.5cm>*{\bullet}="M";
<6cm,-3.5cm>*{\bullet}="N";
<7cm,-2.5cm>*{}="M''";
<7cm,-3.5cm>*{}="N''";
<9.5cm,-2.5cm>*{\bullet}="M'";
<9cm,-3.5cm>*{\bullet}="N'";
<8cm,-2.5cm>*{}="M'''";
<8cm,-3.5cm>*{}="N'''";
"M";"H''" **@{-} ?(0.5)*!/_2mm/{a};
"I''";"N" **@{-} ?(0.2)*!/_2mm/{a'};
"M";"M''" **@{-} ?(0.8)*!/_2mm/{e};
"N";"N''" **@{-} ?(0.8)*!/_3mm/{f};
"M";"N" **@{-} ?(0.5)*!/_2mm/{};
"K''";"M'" **@{-} ?(0.5)*!/_2mm/{b'};
"N'";"L''" **@{-} ?(0.8)*!/_2mm/{b};
"M'''";"M'" **@{-} ?(0.2)*!/_2mm/{e};
"N'''";"N'" **@{-} ?(0.2)*!/_3mm/{f};
"M'";"N'" **@{-} ?(0.5)*!/_2mm/{};
<10cm,-1.5cm>*{\rightarrow};
<11cm,-1cm>*{\bullet}="P";
<11.5cm,-0.5cm>*{\bullet}="Q";
<12.5cm,-0.5cm>*{\bullet}="R";
<13cm,-1cm>*{\bullet}="S";
<13cm,-2cm>*{\bullet}="T";
<12.5cm,-2.5cm>*{\bullet}="U";
<11.5cm,-2.5cm>*{\bullet}="V";
<11cm,-2cm>*{\bullet}="Z";
"P";"Q" **@{-} ?(0.5)*!/_2mm/{};
"Q";"R" **@{-} ?(0.5)*!/_2mm/{};
"R";"S" **@{-} ?(0.5)*!/_2mm/{};
"S";"T" **@{-} ?(0.5)*!/_2mm/{};
"T";"U" **@{-} ?(0.5)*!/_2mm/{};
"U";"V" **@{-} ?(0.5)*!/_2mm/{};
"V";"Z" **@{-} ?(0.5)*!/_2mm/{};
"Z";"P" **@{-} ?(0.5)*!/_2mm/{};
"Q";"V" **@{-} ?(0.5)*!/_2mm/{};
"R";"U" **@{-} ?(0.5)*!/_2mm/{};
"P";"S" **@{-} ?(0.5)*!/_2mm/{};
"Z";"T" **@{-} ?(0.5)*!/_2mm/{};
<5.5cm,-5cm>*{\text{{\bf Algebraic recurrence formula} } 9j \text{(II)}
\,\rightarrow 12j\text{(II)}};
\endxy
\]

\vskip 20pt

The diagrammatic (box--like) and algebraic forms of the
insertion operator to be used recursively for the  generation
of any $3nj$(II) diagram or associated coefficient
are summarized in the following table (where the factor $(2x+1)$
and a phase have been dropped) 
\[
\xy
0*{\mathfrak{I}_{\square}};
<1cm,0cm>*{\leftrightarrow};
<2.5cm,-0.5cm>*{\bullet}="A";
<3.5cm,-0.5cm>*{\bullet}="B";
<3.5cm,0.5cm>*{\bullet}="C";
<2.5cm,0.5cm>*{\bullet}="D";
<2cm,-1cm>*{}="A'";
<4cm,-1cm>*{}="B'";
<4cm,1cm>*{}="C'";
<2cm,1cm>*{}="D'";
"B";"A" **@{-} ?(0.5)*!/_3mm/{f};
"C";"B" **@{-} ?(0.5)*!/_2mm/{c};
"D";"C" **@{-} ?(0.5)*!/_2mm/{e};
"A";"D" **@{-} ?(0.5)*!/_2mm/{d};
"A'";"A" **@{-} ?(0.3)*!/_2mm/{a'};
"B";"B'" **@{-} ?(0.7)*!/_2mm/{b'};
"C";"C'" **@{-} ?(0.7)*!/_2mm/{b};
"D'";"D" **@{-} ?(0.3)*!/_2mm/{a};
<7cm,0cm>*{\leftrightarrow
\, \sum_x
\begin{Bmatrix}
 a & b & x \\
 c & d & e 
 \end{Bmatrix}
 \begin{Bmatrix}
 a' & b' & x \\
 c & d & f 
 \end{Bmatrix}
};
\endxy
\]
Noticing that such an insertion can be applied 
by construction without changing the intrinsic topology
(any prism with basis an $n$--gon is changed into a prism
 with basis an $(n+1)$--gon) the proof of Proposition 2 is completed.  
$\blacksquare$

%%%%%%%%%%%%%%%%%%%%%%%
%%%%%%%%%%%%%               SECTION 3.2
%%%%%%%%%%%%%%%%%%%%%%%%

\subsection{$3nj$(I) diagrams}

Looking at samples of $3nj$(I) diagrams as depicted in \cite{YuLeVa,YuBa},
it is easily inferred that all of them are "cartwheel" configurations,
labeled consistently as shown below
(the arrangement of labels  complies with
the right--hand side of (\ref{RFtypeI}))
\[
\xy
0*{\bullet}="A";
<0.5cm,1cm>*{\bullet}="B";
<1.25cm,1.75cm>*{\bullet}="C";
<2.25cm,2.25cm>*{}="D";
<3.75cm,2.25cm>*{}="E";
<4.75cm,1.75cm>*{}="F";
<5.50cm,1cm>*{\bullet}="G";
<6cm,0cm>*{\bullet}="A'";
<5.5cm,-1cm>*{\bullet}="B'";
<4.75cm,-1.75cm>*{\bullet}="C'";
<3.75cm,-2.25cm>*{}="D'";
<2.25cm,-2.25cm>*{}="E'";
<1.25cm,-1.75cm>*{}="F'";
<0.50cm,-1cm>*{\bullet}="G'";
"A";"B" **@{-} ?(0.5)*!/_3mm/{j_1};
"B";"C" **@{-} ?(0.5)*!/_3mm/{j_2};
"C";"D" **@{-} ?(0.5)*!/_3mm/{j_3};
"F";"E" **@{--} ?(0.5)*!/_3mm/{};
"E";"D" **@{--} ?(0.5)*!/_3mm/{};
"F";"G" **@{-} ?(0.5)*!/_4mm/{j_{n-1}};
"G";"A'" **@{-} ?(0.5)*!/_3mm/{j_n};
"G'";"A" **@{-} ?(0.5)*!/_3mm/{k_n};
"F'";"G'" **@{-} ?(0.5)*!/_3mm/{k_{n-1}};
"F'";"E'" **@{--} ?(0.5)*!/_3mm/{};
"E'";"D'" **@{--} ?(0.5)*!/_3mm/{};
"C'";"D'" **@{-} ?(0.5)*!/_3mm/{k_3};
"B'";"C'" **@{-} ?(0.5)*!/_3mm/{k_2};
"A'";"B'" **@{-} ?(0.5)*!/_3mm/{k_1};
"A";"A'" **@{-} ?(0.1)*!/_3mm/{l_n};
"B";"B'" **@{-} ?(0.1)*!/_3mm/{l_1};
"C";"C'" **@{-} ?(0.1)*!/_3mm/{l_2};
"D";"D'" **@{-} ?(0.1)*!/_3mm/{l_3};
"G'";"G" **@{-} ?(0.1)*!/_3mm/{l_{n-1}};
"F'";"F" **@{-} ?(0.1)*!/_4mm/{l_{n-2}};
<6.5cm,0cm>*{\leftrightarrow};
<9.5cm,0cm>*{
\begin{Bmatrix}
 j_1 & \,& j_2 & \dots & \, & j_n & \,\\
 \, & l_1 & \, & l_2  & \, & \dots & \l_n \\
 k_1 & \, & k_2 & \dots  & \, & k_n &\,
 \end{Bmatrix}};
\endxy
\]

Recall that the computation of the Euler number $\chi$
of a connected configuration of nodes and links drawn in the plane
can be performed by evaluating the quantity $V-E+F$
($V=$ number of vertices, $E=$ number of edges, 
$F=$ number of faces). Here a "face" is any region of the plane
bounded by some polygonal contour with a certain number of vertices,
but graphs with multiple edge--crossings, as those  in the previous picture,
must be suitably rearranged 
 in order to get the correct counting. 
To such a number it must be added $1$ to take in account the 
"external" region, namely the portion of the  plane not included
into the graph. 
\vskip 12pt

\noindent {\bf Proposition 3}.\\ 
{\em Any $3nj$ diagram of type I is 
a graph with Euler number $\chi =1$.}

\vskip 12pt 
The proof proceed by induction on $n$ and will be completed
after Proposition 4. According to
the definition given in the introduction to this section,
it is worth noting preliminary that an Euler number
equal to 1 does not correspond to any embedding
of graphs on an {\em oriented}
surface. Actually the definition of $\chi$ can be generalized to
(embeddings of graphs on) {\em nonorientable} closed surfaces $\tilde{\Sigma}$
by introducing the so--called "crosscap number" $c$, namely the number
of M\"obius bands one must attach to the sphere $S^2$ to obtain
the desired surface. Such a "nonorientable genus"
is related to the Euler number of $\tilde{\Sigma}_c$
by $\chi( \tilde{\Sigma}_c) = (2 - c)$ and $\chi=1$ corresponds
to the (minimum) crosscap number $c=1$ characterizing  
the real projective plane $P$ ($\mathbb{RP}^2$ in the standard
mathematical notation).

The $9j$(I) diagram was explicitly shown to be embeddable in $P$ in 
\cite{PoRe}, but this feature is actually well known because of 
its  isomorphism with the
complete bipartite graph $K_{3,3}\,$ (see
{\em e.g.} \cite{GrTu})
\footnote{At the end of section 2.1 the
embedding of $9j$(I) into the orientable, genus--1 torus $T$ was considered.
The fact that $K_{3,3}\,$ has the same nonorientable and orientable genus
is exceptional: most graphs have different values for $g$ and $c$.}.  \\
Taken for granted this result, the following Proposition
deals with the existence of a well--defined
insertion operation characterized by the the fact that
it preserves the Euler number, namely the variation 
$\Delta_{n-1,n} \,\chi$
of $\chi$ in passing from $3(j-1)$(I) to  $3nj$(I)
diagrams is zero.

\vskip 12pt

\noindent {\bf Proposition 4}.\\
{\em There exists (up to symmetry) one kind
of "twisted" insertion operation, denoted by
 $\mathfrak{I}_{\bowtie}\,$, that generates
the $3nj$(I) from the $3(n-1)j$(I) for
any $n\geq 3$ according to the recurrence formula
(\ref{RFtypeI}). Moreover $\mathfrak{I}_{\bowtie}\,$
is topology--preserving with respect to embeddings
in $P$.}

The proof can be carried out by resorting  to the
graphical method as already done in section 3.1 for type II diagrams. 
 In the figure below  the generation of the $3nj$(I)
from the $3(n-1)j$(I) 
according to the algebraic recurrence formula (\ref{RFtypeI}) is shown.
The summation
over $x$ implies  the cancellation of both edges labeled by $x$
and nodes which lose an incident edge; the last step would consist
in joining back edges with the same labels 
to get the $3nj$(I) diagram represented at the beginning of
this section. Note that, up to relabeling, the same construction
could be performed on anyone of the $(n-1)$ rays of the $3(n-1)j$(I) 
and the resulting coefficients have the same values owing 
to their symmetry properties \cite{YuLeVa}.\\
\[
\xy
0*{};
<0.5cm,0cm>*{\sum_x};
<2cm,0cm>*{\bullet}="A";
<3cm,1cm>*{\bullet}="B";
<4cm,0cm>*{\bullet}="C";
<3cm,-1cm>*{\bullet}="D";
<5cm,0cm>*{\bullet}="E";
<5.5cm,1cm>*{\bullet}="F";
<5.5cm,-1cm>*{\bullet}="G";
<6.25cm,1.75cm>*{}="H";
<7.75cm,1.75cm>*{}="I";
<9cm,0cm>*{\bullet}="E'";
<8.5cm,-1cm>*{\bullet}="F'";
<8.5cm,1cm>*{\bullet}="G'";
<7.75cm,-1.75cm>*{}="H'";
<6.25cm,-1.75cm>*{}="I'";
<10cm,0cm>*{\bullet}="A'";
<11cm,1cm>*{\bullet}="B'";
<12cm,0cm>*{\bullet}="C'";
<11cm,-1cm>*{\bullet}="D'";
"A";"B" **@{-} ?(0.5)*!/_3mm/{l_n};
"B";"C" **@{-} ?(0.5)*!/_3mm/{j_1};
"C";"D" **@{-} ?(0.5)*!/_4mm/{k_{n-1}};
"D";"A" **@{-} ?(0.5)*!/_3mm/{l_{n-1}};
"A";"C" **@{-} ?(0.3)*!/_2mm/{x};
"D";"B" **@{-} ?(0.3)*!/_2mm/{k_n};
"E";"F" **@{-} ?(0.5)*!/_2mm/{j_1};
"F";"H" **@{-} ?(0.5)*!/_2mm/{};
"H";"I" **@{--} ?(0.5)*!/_2mm/{};
"I";"G'" **@{-} ?(0.5)*!/_2mm/{};
"G'";"E'" **@{-} ?(0.5)*!/_4mm/{j_{n-1}};
"E'";"F'" **@{-} ?(0.5)*!/_2mm/{k_1};
"F'";"H'" **@{-} ?(0.5)*!/_2mm/{};
"H'";"I'" **@{--} ?(0.5)*!/_2mm/{};
"I'";"G" **@{-} ?(0.5)*!/_2mm/{};
"G";"E" **@{-} ?(0.5)*!/_4mm/{k_{n-1}};
"E";"E'" **@{-} ?(0.2)*!/_2mm/{x};
"F";"F'" **@{-} ?(0.2)*!/_3mm/{l_1};
"G'";"G" **@{-} ?(0.8)*!/_4mm/{l_{n-1}};
"A'";"B'" **@{-} ?(0.5)*!/_4mm/{j_{n-1}};
"B'";"C'" **@{-} ?(0.5)*!/_4mm/{l_{n-1}};
"C'";"D'" **@{-} ?(0.5)*!/_4mm/{l_n};
"D'";"A'" **@{-} ?(0.5)*!/_3mm/{k_1};
"A'";"C'" **@{-} ?(0.3)*!/_2mm/{x};
"D'";"B'" **@{-} ?(0.3)*!/_2mm/{j_n};
\endxy
\]
%%SECOND PART
\[
\xy
0*{};
<0.7cm,0cm>*{=};
<2cm,0.5cm>*{}="A";
<3cm,1cm>*{\bullet}="B";
<4cm,0.5cm>*{}="C";
<4cm,-0.5cm>*{}="D";
<3cm,-1cm>*{\bullet}="E";
<2cm,-0.5cm>*{}="F";
<5cm,0.5cm>*{}="G";
<6cm,1cm>*{\bullet}="H";
<7cm,1.5cm>*{}="I";
<8cm,1.5cm>*{}="L";
<9cm,1cm>*{\bullet}="M";
<10cm,0.5cm>*{}="N";
<5cm,-0.5cm>*{}="O";
<6cm,-1cm>*{\bullet}="P";
<7cm,-1.5cm>*{}="Q";
<8cm,-1.5cm>*{}="R";
<9cm,-1cm>*{\bullet}="S";
<10cm,-0.5cm>*{}="T";
<11cm,0.5cm>*{}="A'";
<12cm,1cm>*{\bullet}="B'";
<13cm,0.5cm>*{}="C'";
<13cm,-0.5cm>*{}="D'";
<12cm,-1cm>*{\bullet}="E'";
<11cm,-0.5cm>*{}="F'";
"A";"B" **@{-} ?(0.5)*!/_3mm/{l_n};
"B";"C" **@{-} ?(0.5)*!/_3mm/{j_1};
"D";"E" **@{-} ?(0.5)*!/_4mm/{k_{n-1}};
"E";"F" **@{-} ?(0.5)*!/_3mm/{l_{n-1}};
"E";"B" **@{-} ?(0.5)*!/_2mm/{k_n};
"G";"H" **@{-} ?(0.5)*!/_3mm/{j_1};
"H";"I" **@{-} ?(0.5)*!/_3mm/{};
"I";"L" **@{--} ?(0.5)*!/_3mm/{};
"L";"M" **@{-} ?(0.5)*!/_3mm/{};
"M";"N" **@{-} ?(0.5)*!/_4mm/{j_{n-1}};
"T";"S" **@{-} ?(0.5)*!/_3mm/{k_1};
"S";"R" **@{-} ?(0.5)*!/_3mm/{};
"Q";"R" **@{--} ?(0.5)*!/_3mm/{};
"Q";"P" **@{-} ?(0.5)*!/_3mm/{};
"P";"O" **@{-} ?(0.5)*!/_3mm/{k_{n-1}};
"H";"S" **@{-} ?(0.3)*!/_3mm/{l_1};
"M";"P" **@{-} ?(0.7)*!/_4mm/{l_{n-1}};
"A'";"B'" **@{-} ?(0.5)*!/_4mm/{j_{n-1}};
"B'";"C'" **@{-} ?(0.5)*!/_4mm/{l_{n-1}};
"D'";"E'" **@{-} ?(0.5)*!/_4mm/{l_n};
"E'";"F'" **@{-} ?(0.5)*!/_3mm/{k_1};
"B'";"E'" **@{-} ?(0.5)*!/_3mm/{k_n};
\endxy
\]
The diagrammatic [algebraic] form of the twisted
insertion operator $\mathfrak{I}_{\bowtie}\,$
to be used recursively for the  generation
of any $3nj$(I) diagram [coefficient]
is summarized in the following table (where the factor $(2x+1)$
and a phase have been dropped) 
%TWISTED INSERTION OPERATOR
\[
\xy
0*{\mathfrak{I}_{\bowtie}};
<1cm,0cm>*{\leftrightarrow};
<2cm,-0.5cm>*{\bullet}="A";
<3.5cm,-0.5cm>*{\bullet}="B";
<3.5cm,0.5cm>*{\bullet}="C";
<2cm,0.5cm>*{\bullet}="D";
<1.5cm,-1cm>*{}="A'";
<4cm,-1cm>*{}="B'";
<4cm,1cm>*{}="C'";
<1.5cm,1cm>*{}="D'";
"A";"A'" **@{-} ?(0.3)*!/_4mm/{k_{n-1}};
"B'";"B" **@{-} ?(0.3)*!/_2mm/{k_1};
"C";"C'" **@{-} ?(0.7)*!/_4mm/{j_{n-1}};
"D'";"D" **@{-} ?(0.3)*!/_3mm/{j_1};
"A";"D" **@{-} ?(0.5)*!/_2mm/{k_n};
"C";"B" **@{-} ?(0.5)*!/_2mm/{j_n};
"C";"A" **@{-} ?(0.7)*!/_3mm/{l_{n-1}};
"D";"B" **@{-} ?(0.2)*!/_2mm/{l_n};
<8cm,0cm>*{\leftrightarrow
\, \sum_x
\begin{Bmatrix}
 j_1 & k_{n-1} & x \\
 l_{n-1} & l_n & k_n 
 \end{Bmatrix}
 \begin{Bmatrix}
k_1 & j_{n-1} & x \\
 l_{n-1} & l_n & j_n 
 \end{Bmatrix}
};
\endxy
\]
The topological invariance of such an operation is easily proved
by observing that the variation of the Euler number
owing to one insertion  is given by
\begin{equation}\label{deltachi1}
\Delta_{n-1,n}\, \chi \, [\mathfrak{I}_{\bowtie}\,]\,=\,
\Delta V - \Delta E + \Delta F \,=\, 
2-3+1 \equiv 0
\end{equation}
where $\Delta V$ is the number of new
vertices, $\Delta E$  the number of new
edges and $\Delta F$ the number of new faces.
(Recall  that any type of $3nj$ diagram has $2n$ vertices and 
$3n$ edges and note that the $(n+1)$ "faces" for type I
can by quickly selected by looking at the $n$ 
cycles that include each of the $n$ rays and adding $+1$
for the external region.)

Now the proof of Proposition 3 can be completed by induction.
Given that $\chi \, [9j]=1$ and assuming
$\chi\, [3(n-1)j]=1$, by (\ref{deltachi1}) we
get
\begin{equation} \label{deltachi2}
\chi \, [3nj] \,=\, \chi [3(n-1)j] \,+\,\Delta_{n-1,n}\,
\chi \, [\mathfrak{I}_{\bowtie}\,]\,=\,1
\end{equation}
for any $n \geq 3$.
$\blacksquare$

A comment is in order about the possibility of generating 
recursively the other types of $3nj$ diagrams arising for $n >4$
\cite{YuLeVa,YuBa}. We argue that, on applying suitable sequences of
$N$ insertion operations $\mathfrak{I}_{\square\,}$  (Prop. 2)
and $\mathfrak{I}_{\bowtie\,}$ (Prop. 4) starting from the simplest
graph --namely the complete quadrilateral associated with the
$6j$-- {\em all types} of $3(2+N)j$ diagrams can be generated.
Actually the topological invariance 
proved in the two propositions holds true only 
when $\mathfrak{I}_{\square\,}$ is applied to a type II diagram
(and similarly $\mathfrak{I}_{\bowtie\,}$ to a type I diagram)
under the circumstances described above. However, different
positions of the insertions and/or actions on the other type
of diagrams would provide a variety of configurations with possible
changes of the underlying topology.
Work is in progress in this direction.

In view of applications to be addressed 
in the following we conclude this section by noticing 
that every $3nj$(I) diagram has a Hamiltonian circuit 
of length $2n$, as the samples of $12j, 15j, 18j$(I)
shown below make it manifest. The Hamiltonian cycle would be always consistently
labeled by the ordered set $j_1,j_2,\dots,j_n,$  $k_1,k_2,$ $\dots,k_n$ and
the rays by $l_1,l_2,\dots,l_n$. 
%%%%%%%%%%%%%%%%%HAMILT CYCLES TYPE I
\[
\xy
/r18mm/: (0,0), {\xypolygon8{~<{-}~>{*}
{}}},(3,0),
{\xypolygon10{~<{-}~>{*}
{}}}
,(6,0),
{\xypolygon12{~<{-}~>{*}
{}}}
\endxy
\]
Other non--Hamiltonian cycles
appear as well in these diagrams. 
There are both cycles of
length equal to 
$n+1$ (including one ray and $n$ edges on the perimeter) 
as well as cycles of length equal to the girth number, which is
$4$ for each $n$ as already happened for   
$3nj$(II) diagrams (now these 4--cycles 
are associated with "twisted" configurations
of four edges labeled cyclically by $k_i,j_i, l_{i-1},
l_i$).

%%%%%%%%%%%%%%%%%%%%%%%%%
%%%%%%%%%%%%%%%%%%%%%%%%
%%%%%%%%                     SECTION 4: INTRODUCTORY PART
%%%%%%%%%%%%%%%%%%%%%%%%%%

\section{Asymptotic disentangling
  of 3nj diagrams}
  
In two previous papers \cite{ABFMR1,ABFMR2}   the mathematical apparatus of
angular momenta recoupling theory
\cite{BiLo9} (topic 12) has been shown to provide a unifying background structure
for a number of phenomena and applications ranging from atomic, molecular
and nuclear physics to quantum computing and gravity
({\em cfr.} also \cite{CaMaRa}).
The term   {\em spin networks}, originally introduced by Penrose \cite{Pen}
in connection with a model combinatorial spacetime, was
used as a synonym of $3nj$ diagrams, namely connected
and closed regular graphs whose $2n$ nodes are associated with 
triads of angular momentum variables, each satisfying
triangle inequality.
By {\em semiclassical} spin networks we mean, as in  \cite{ABFMR1,ABFMR2}, those families
of $3nj$ diagrams that arise when all or just a few of the
$3n$ spin variables become large (namely the 
angular momentum labels attached to them 
are $\gg 1$ in $\hbar$ units). The issue of semiclassical analysis
of angular momentum quantum transition amplitudes dates back
to the beginning of quantum mechanics and  
we do not insist here on the wide range
of applications of such techniques. 
Rather, we are going
to reach a general assessment starting from what has been 
done for the $9j$ symbol
in \cite{AnAqFe}, where the case in which six edges
become large was shown to provide the {\em disentangling}
of the underlying network, in the sense explained in section 4.1 below.
(Note that a more formal approach based on geometric quantization
techniques for both $3j$ (Wigner) coefficient \cite{AqHaLi2}
and  $6j$ symbol \cite{Ragni}
can be carried out as well.)

Whatever kind of asymptotics we want to address, the first
object to be analyzed is the 3--valent node itself, namely the
basic building block of any $3nj$ diagram. Given that
a node is naturally associated with a $3j$ (Wigner) coefficient,
and forgetting its dependence on magnetic quantum numbers,
the possible choices of small/large angular momentum values
are summarized in the following picture, where solid lines
and small letters stand for truly "quantum" labels while
capital letters and double dotted lines represent "semiclassical"
large values.

\[
\xy
0*{\text{(quantum}\, 3j)};
<1cm,1cm>*{}="A";
<3cm,1cm>*{}="B";
<2cm,-1.4cm>*{}="C";
<2cm,0cm>*{\bullet}="D";
<5cm,1cm>*{}="A'";
<7cm,1cm>*{}="B'";
<6cm,-1.4cm>*{}="C'";
<6cm,0cm>*{\bullet}="D'";
<9cm,1cm>*{}="A''";
<11cm,1cm>*{}="B''";
<10cm,-1.4cm>*{}="C''";
<10cm,0cm>*{\bullet}="D''";
"A";"D" **@{-} ?(0.3)*!/_2mm/{a};
"D";"C" **@{-} ?(0.5)*!/_2mm/{c};
"D";"B" **@{-} ?(0.7)*!/_2mm/{b};
"A'";"D'" **@{:} ?(0.3)*!/_3mm/{A};
"D'";"C'" **@{-} ?(0.5)*!/_2mm/{c};
"D'";"B'" **@{:} ?(0.7)*!/_3mm/{B};
"A''";"D''" **@{:} ?(0.3)*!/_3mm/{A};
"D''";"C''" **@{:} ?(0.5)*!/_3mm/{C};
"D''";"B''" **@{:} ?(0.7)*!/_3mm/{B};
<2cm,-2cm>*{\mathbf{(0,3) \; case}};
<6cm,-2cm>*{\mathbf{(2,1) \; case}};
<10cm,-2cm>*{\mathbf{(3,0) \;case}};
\endxy
\]
Here we recover graphically the well known admissible
asymptotics for $3j$ coefficients, that is 
"if one spin value is large then
at least one of the other incident spins must be large as well".
Moreover, if the right--hand case $(3,0)$ occurs at anyone of the
$2n$ node of a $3nj$ diagram, the resulting configuration
turns out to be related to an amplitude whose distribution
represents (a suitable number of) decoupled harmonic oscillators.
Thus the "quantum entanglement of states", measured
precisely by the transition amplitudes associated with $3nj$ coefficients,
is lost when approaching the classical limit, namely when 
the underlying quantum system undergoes a complete decoherence.
     
The prototype of this phenomenon  was addressed in the famous paper
by Ponzano and Regge \cite{PoRe} on "Semiclassical limit of Racah coefficients"
dealing with the simplest spin network, the $6j$ symbol itself.
In what follows we are going to explore and classify  
asymptotic regimes in
which different kinds of "asymptotic disentangling"
arise in configurations that include  
$(2,1)$--nodes.
(Recall that the well--known formulas for the $(3,3)$
and $(4,2)$ $6j$ cases, corresponding to the cycles shown in section
2.1, are listed as Eqns. 2 and 3 in \cite{AnAqFe}).

As a case study  we review first
the main results on the $9j$(I) proved in \cite{AnAqFe}
by supporting them with new numerical simulations.

%\vfill
%\newpage
%%%%%%%%%%%%%%%%%%%%%%%%%%%%%
%%%%%%%%%%%%%%                   SECTION 4.1
%%%%%%%%%%%%%%%%%%%%%%%%%%%%

\subsection{The 9j case}

 The focus of this section is on the $9j$(I) symbol, 
well known in atomic spectroscopy for its role as the 
matrix element of the transformation between $LS$ and $jj$ 
coupling schemes, but exhibiting features prototypical 
of more complex spin networks.  In quantum chemistry, 
it appears for example in matrix elements when Sturmian 
or momentum space orbitals are employed \cite{AqCaAv}.
On the basis of the remarks
at the end of section 2.2, 
it is worth noting that the physical situations where some of the angular
 momenta become large,  correspond mathematically
to the case of discrete
 functions becoming continuous in the limit, as we are going to
check in what follows.  
Note however that we are not going to
 address the analog of Ponzano--Regge asymptotics ($9$ entries
 of the $9j$ large, or $(9,0)$ expansion) and refer the reader to
 \cite{Russi} Sec. 10.7 for details (the same reference
is the basic one for notations and algebraic expressions used
in this section). 

All cases where some entries are large and some small can
 be reduced to the
$(4,5)$ or  "large corner" case and to the  $(6,3)$ or   
"small diagonal" case
by permutation symmetries. The associated Yutsis graphs
with dotted edges representing large entries are
\[
\xy
0*{\bullet}="A";
<-1cm,1.7cm>*{\bullet}="B";
<0cm,3.4cm>*{\bullet}="C";
<2cm,3.4cm>*{\bullet}="D";
<3cm,1.7cm>*{\bullet}="E";
<2cm,0cm>*{\bullet}="F";
"A";"B" **@{-} ?(0.5)*!/_2mm/{};
"B";"C" **@{:} ?(0.5)*!/_2mm/{};
"C";"D" **@{:} ?(0.5)*!/_2mm/{};
"D";"E" **@{:} ?(0.5)*!/_2mm/{};
"E";"F" **@{-} ?(0.5)*!/_2mm/{};
"F";"A" **@{-} ?(0.5)*!/_2mm/{};
"A";"D" **@{-} ?(0.3)*!/_2mm/{};
"C";"F" **@{-} ?(0.3)*!/_2mm/{};
"B";"E" **@{:} ?(0.7)*!/_2mm/{};
<6cm,0cm>*{\bullet}="A'";
<5cm,1.7cm>*{\bullet}="B'";
<6cm,3.4cm>*{\bullet}="C'";
<8cm,3.4cm>*{\bullet}="D'";
<9cm,1.7cm>*{\bullet}="E'";
<8cm,0cm>*{\bullet}="F'";
"A'";"B'" **@{:} ?(0.5)*!/_2mm/{};
"B'";"C'" **@{:} ?(0.5)*!/_2mm/{};
"C'";"D'" **@{:} ?(0.5)*!/_2mm/{};
"D'";"E'" **@{:} ?(0.5)*!/_2mm/{};
"E'";"F'" **@{:} ?(0.5)*!/_2mm/{};
"F'";"A'" **@{:} ?(0.5)*!/_2mm/{};
"A'";"D'" **@{-} ?(0.3)*!/_2mm/{};
"C'";"F'" **@{-} ?(0.3)*!/_2mm/{};
"B'";"E'" **@{-} ?(0.7)*!/_2mm/{};
<1cm,-1cm>*{\mathbf{(4,5) \; case}};
<7cm,-1cm>*{\mathbf{(6,3) \;case}};
\endxy
\]
Recall that the basic expression of the $9j$ as a sum over $x$
of the product of three $6j$ was already given in (\ref{9jI}).
By slightly changing spin labels to comply with notations 
in \cite{AnAqFe}, it is rewritten as 
\begin{equation}\label{9jIasym}
\sum_x (-)^{2x}(2x+1)
 \begin{Bmatrix}
 a  & b & c \\
 f  & i  & x 
 \end{Bmatrix} 
 \begin{Bmatrix}
 d  & e & f \\
 b  & x  & h
 \end{Bmatrix} 
 \begin{Bmatrix}
 g  & h & i \\
 x & a  & d 
 \end{Bmatrix}\;=\; 
 \begin{Bmatrix}
 a  & b & c \\
 d  & e  & f \\
g & h & i 
 \end{Bmatrix}, 
\end{equation}
where $\text{max} \{ |a-i|,|b-f|,|h-d| \}
\leq x \leq \text{min} \{ a+i,b+f,h+d \}$.\\
The $(6,3)$ asymptotic case is derived from this
defining equation
by suitably applying the well known $(3,3)$ expansions
for the $6j$ \cite{PoRe,ScGo} and the final result
reads (recall that capital letters stand for large entries)
\begin{equation}\label{9j63}
 \begin{Bmatrix}
 a & B & C \\
 D & e & F \\
 G & H & i 
 \end{Bmatrix}
 \xrightarrow{(6,3)}
 \;\frac{\;\;(-)^{B+H-e}}{\Gamma_1\,\Gamma_2}\, 
 \mathbf{d}^{\,a}_{C-B,G-D\,}\,(\cos \theta )\;
  \mathbf{d}^{\,i}_{F-C,H-G\,}\,(\cos \theta ),
 \end{equation}
where there appear two Jacobi polynomials (Wigner $\mathbf{d}$--matrices
of a cosine),
with principal quantum numbers $a$, $i$ 
(two of the "small" entries) and
argument given by
$$ 
\cos \theta\,=\, \frac{(2e+1)^2 - (B+F+1)^2 - (D+H+1)^2}
{2(B+F+1)(D+H+1)}.
$$
Note that the role of magnetic quantum numbers in $\mathbf{d}^j_{MM'}$
is played here by differences of "large" spin variables which vary
consistently between $-j$ and $+j$ in integer steps.
Finally, the multiplicative factors $\Gamma$ read
$$
\Gamma_1\,=\, \left\{ [\tfrac{2}{3}(B+C+f)+1][\tfrac{2}{3}(D+G+H)+1]
\right\}^{1/2} ;
$$
$$
\Gamma_2\,=\,[(B+F+1)(D+H+1)]^{1/2}.
$$
The $(4,5)$ case was addressed years ago by Watson \cite{Wat} and his
results are summarized in 
 \begin{equation}\label{9j45}
 \begin{Bmatrix}
 A & b & C \\
 d & e & f \\
 G & h & I 
 \end{Bmatrix}
 \xrightarrow{(4,5)}
 \,\frac{\;\;(-)^{b+C-d-G}}{2X+1} 
 \begin{pmatrix}
 b & e & h \\
 C-A & \epsilon & G-I
\end{pmatrix}
\begin{pmatrix}
 d & e & f \\
 G-A & \epsilon & C-I
\end{pmatrix}
 \end{equation}
 where there appear two Wigner $3j$ symbols and $X=\tfrac{1}{4}(A+C+G+I)$,
 $\epsilon = A-C-G+I$.
 
Both formulas can then be taken as an illustration of  "disentangling" of 
the associated $9j$ networks because in the above 
semiclassical limits  no summation ("superposition") appears anymore.
Both $\mathbf{d}$--matrices and Clebsch--Gordan (Wigner) coefficients
 represent well defined classes of orthogonal 
polynomials belonging to the Askey hierarchy
as  illustrated for instance in \cite{Ragni} in  connection
with the quantum theory of angular momentum. 
As is well known, the issue of separability of variables in quantum mechanics 
is associated, by group 
theoretical arguments, to continuous or discrete symmetries, 
and allows the expression 
of  wavefunctions of a composite
system as a product of wavefunctions of  subsystems.  
On the other hand, nonseparability requires introduction of coupling by expanding 
on a linear combination of a set of such basis wavefunctions.
Interpreting the original expression of the $9j$ given in (\ref{9jI})
(or in (\ref{9jIasym}))
as an expansion of  a global "spin network"
wavefunction, where the entries are discrete variables, 
the asymptotics (\ref{9j63}) and (\ref{9j45})
actually lead to wavefunctions proper of  separated systems
(note that the surviving angular momentum functions depend on both truly
quantum discrete variables and on semiclassical, continuous ones).
Indeed,  from the analysis carried out in section  2.1, we see that the variables
are explicitly designated by $c,d$ and $h$ in the discrete case of a coupled set of
difference equations, and by $x,y,z$ in the continuous limit of a coupled set of partial
differential equations. This appears to have been anticipated in an early
remark by Neville \cite{Nev}.

From the graphical viewpoint, what is left
after  "deletion" of the dotted edges
--associated to real--valued variables that tend to infinity-- is in both cases  a graph 
with no more closed cycles and thus "trivial".
They look like isolated edges (case $(6,3)$) or tree configurations (case $(4,5)$), 
which are both topologically "contractible". 
The "entangled" nature of higher $3nj$ cases can be discussed along 
similar lines and according to remark (ii) of section 2.2.
In the next section we are going to deal systematically with their disentangling. 
Note finally that such disentangling phenomena might be discussed and 
interpreted from many other viewpoints  
since spin network graphs ($3nj$ diagrams) are encountered in
 quantum gravity models, discrete mathematics, and quantum computing
(\cite{ABFMR1,ABFMR2,CaMaRa} and reference therein). We plan to investigate new applications
of our asymptotic techniques in these contexts. 
 
 \vskip 5pt
 
Further important insight on disentangling  may come from computational studies.  
In \cite{AnAqFe} Wigner $\mathbf{d}$--functions, 
Clebsch--Gordan coefficients  and other special functions 
were calculated by directly 
summing their defining series using multiprecision arithmetic (MPFUN90)
\cite{MPFUN}.  
The multiple precision arithmetic allows convenient calculation 
of hypergeometric functions, 
of small and large argument by their series definition, and without 
the necessity of using recurence relations, 
integral and rational representations, or asymptotic approximations.  
On the basis of defininig relations
and formulas for the functions given in various chapters of \cite{Russi}
the basic problem encountered in these summations is that 
some of the terms in the sum may become very large 
($\sim 10^{100}$ or larger).
 However the sum must be a number in a small finite range, 
say $[-1,1]$, and normal computer precision (8 to 32 decimal places)
 will not generally produce accurate summation. 
On the other hand, multiple precision arithmetic packages that 
allow calculations with thousands (or millions!) of digits sum the 
hypergeometric series easily for a spin variable $j$ of magnitude 
much larger than one thousand.
The asymptotic formulas  (\ref{9j63}) and (\ref{9j45}) 
have been already tested in \cite{AnAqFe} 
by calculating  $9j$ symbols and 
their approximations for huge 
randomly selected sets of  $j$ values but 
we conclude this section by presenting
new numerical experiments, illustrated in the following
and displayed in the figures below.

The large entries on the coefficients 
(either 6 or 4 according to either (\ref{9j63}) or (\ref{9j45})) 
have values ranging from a given abscissa, 
$J$  to  $J + j_m$, where $J$ is chosen between 20 to 100.  The small  $j$  
(integer and half integer values) are randomly selected from the range  
$[0 ,  j_m]$.   The calculations use $1000$ nonzero $9j$ values for each $J$ and 
integer and half integer small $j$ for $j_m = 2, 9/2, 19/2, 29/2$.  
We calculate the {\em rms deviations}, {\em i.e.} the root mean square deviations 
of the approximate $9j$ from the exact values, and the {\em rms magnitudes} of 
the $9j$ symbols.  
We take the ratio of these quantities to evaluate the 
{\em fractional error} in the disentangled $9j$ symbol.  
Then the numerical experiments 
show that the fractional error decreases as $J^{-1}$  for both Equations 
(\ref{9j63}) and (\ref{9j45})) 
while the rms magnitudes decrease as $J^{-2}$  for Eq.(\ref{9j63}) 
and  $J^{-1}$  
for Eq.(\ref{9j45})). 
 
Experiments with different values for $j_m$ show that the 
fractional errors for both equations scale as  $j^2_m/J$.  For the six large 
angular momenta formula, the fractional error is approximately $0.3 j^2_m/J$.  
For the four large angular momenta formula, the  fractional error  
is approximately  $0.2 j^2_m/J$.    
The Watson equation  (\ref{9j45}) produces a significant 
number of zero values (15 to 20 \%) for small but nonzero exact $9j$ symbols.  
These zeros result from combinations of angular momenta that yield zero values 
for the $3j$ symbols.  Preliminary numerical checks indicate a correlation with 
the previously mentioned phenomenon of the appearance of nodes in separable 
conditions.  Remarkably the rms magnitudes of the $9j$ symbols that correspond 
to zero approximate values scale as  $J^{-2}$  and have approximately the same values 
as the rms deviation values.  Hence the quality of the overall  approximation
 not degraded even if the cases with zero approximate values are included.

\begin{figure*}[htbp]
	\centering
		\includegraphics[width=0.50\textwidth]{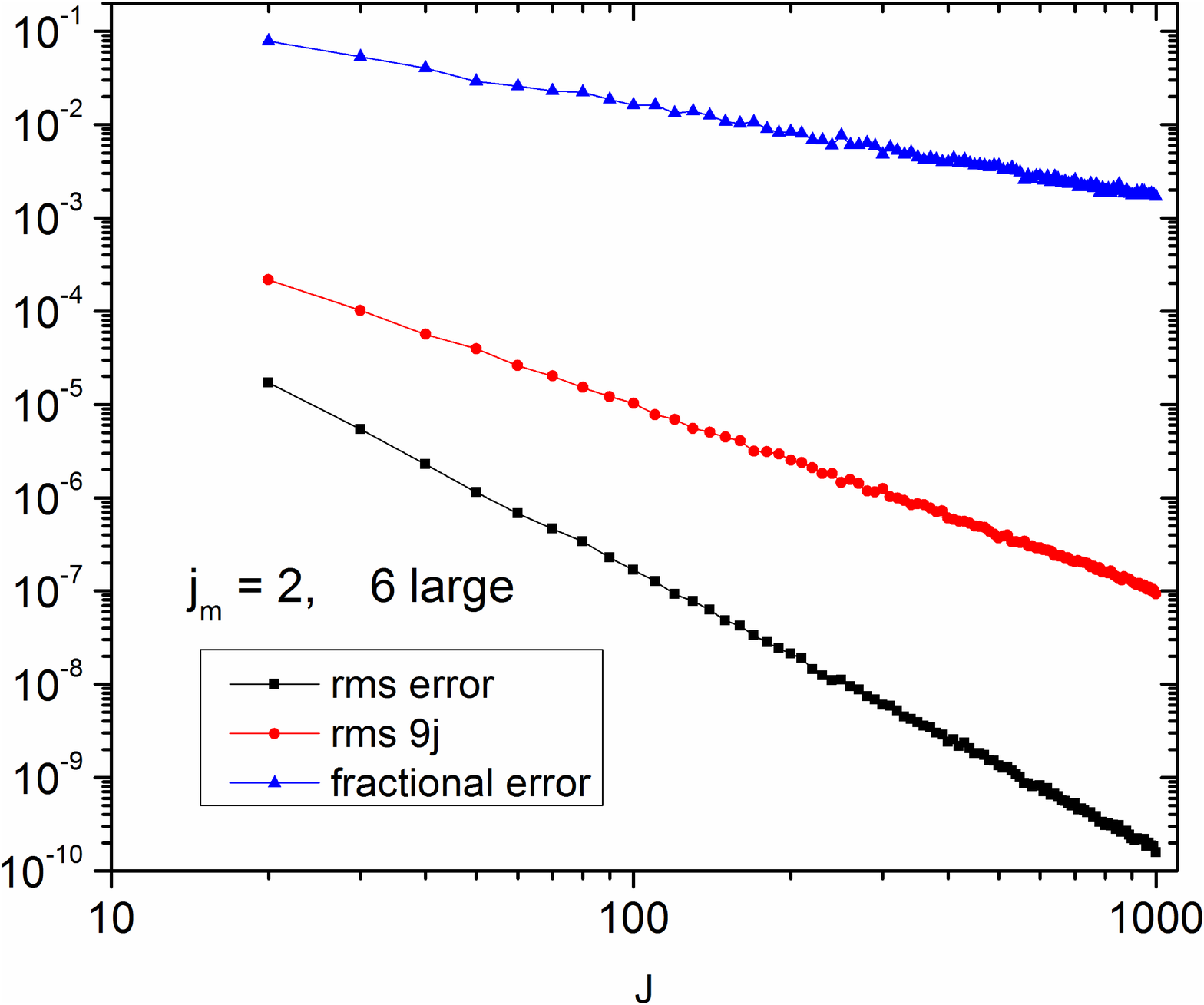}
		\caption{}
	\label{Fig1}
\end{figure*}

\vfill
\newpage

 \begin{figure*}[htbp]
	\centering
		\includegraphics[width=0.50\textwidth]{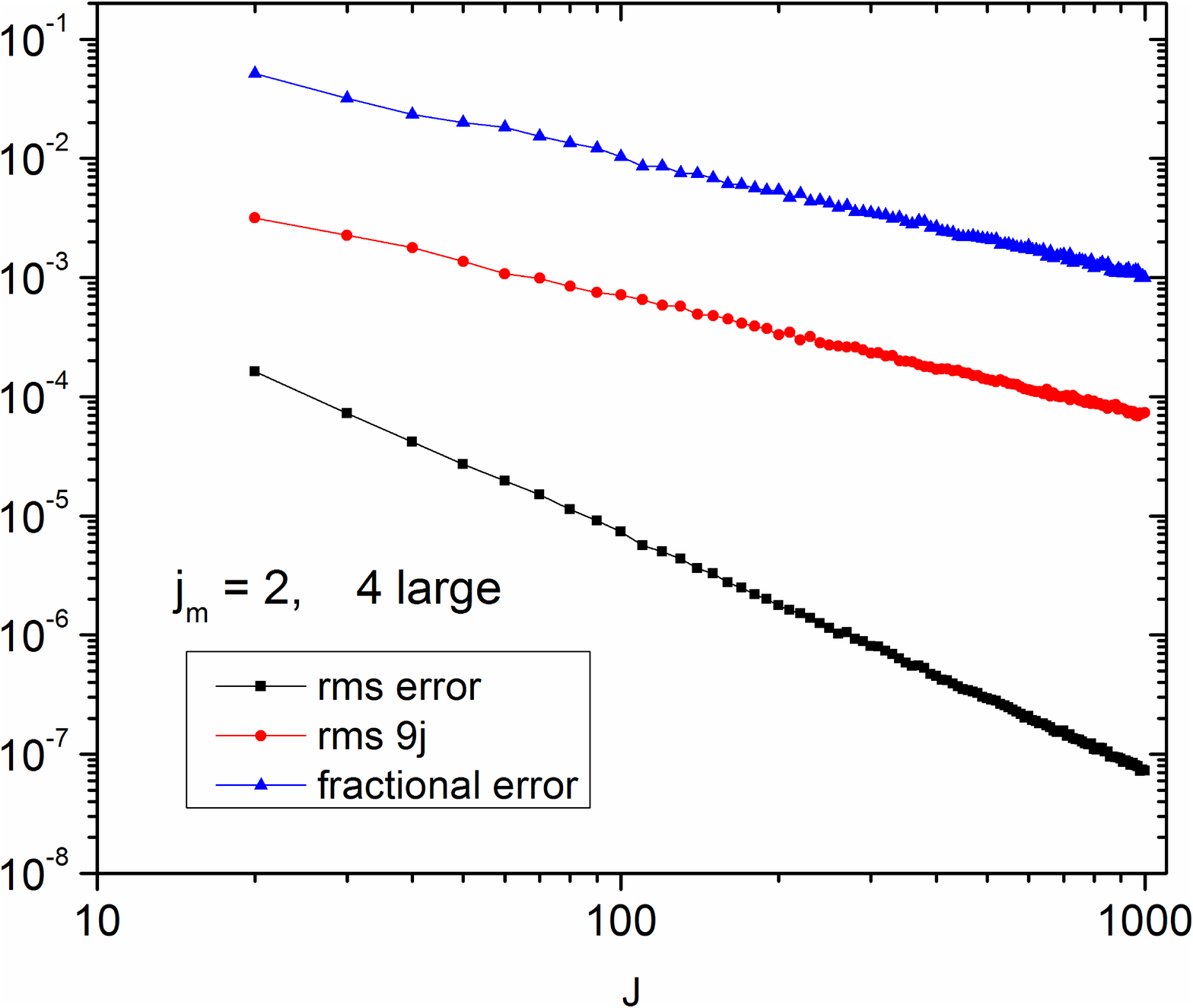}
		\caption{}
	\label{Fig2}
\end{figure*}

\begin{figure*}[htbp]
	\centering
		\includegraphics[width=0.50\textwidth]{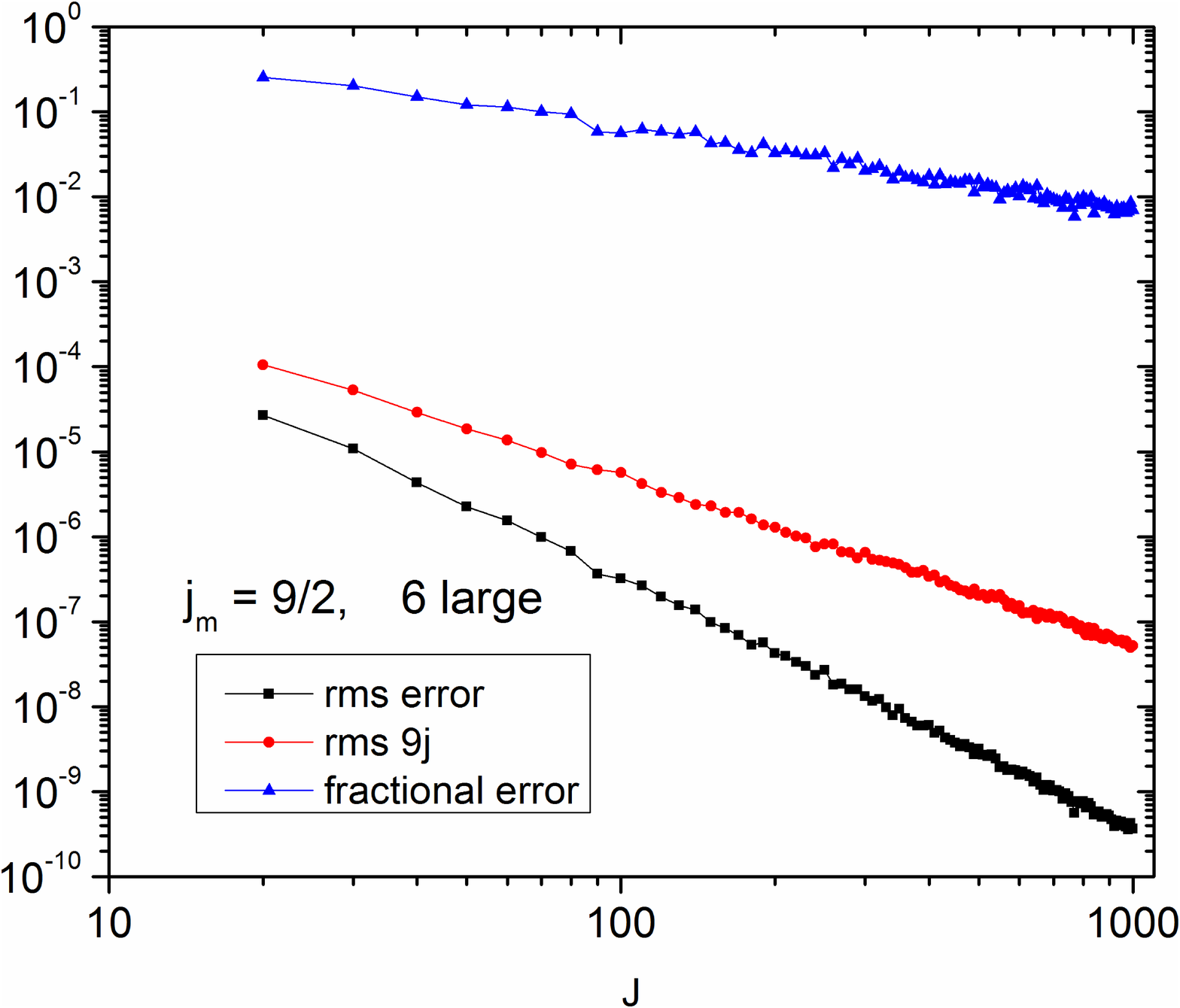}
		\caption{}
	\label{Fig3}
\end{figure*}

\begin{figure*}[htbp]
	\centering
		\includegraphics[width=0.50\textwidth]{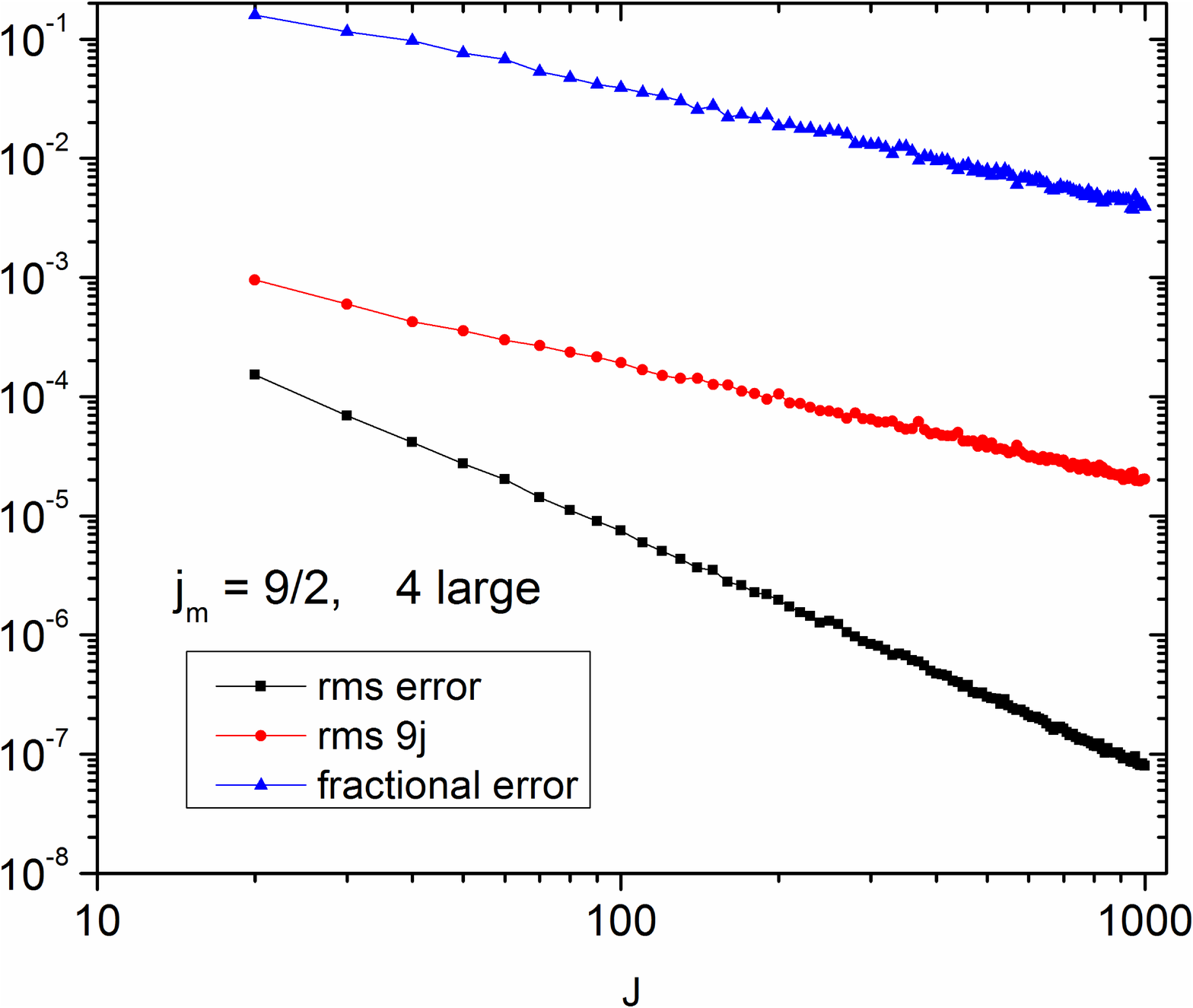}
		\caption{}
	\label{Fig4}
\end{figure*}

\begin{figure*}[htbp]
	\centering
		\includegraphics[width=0.50\textwidth]{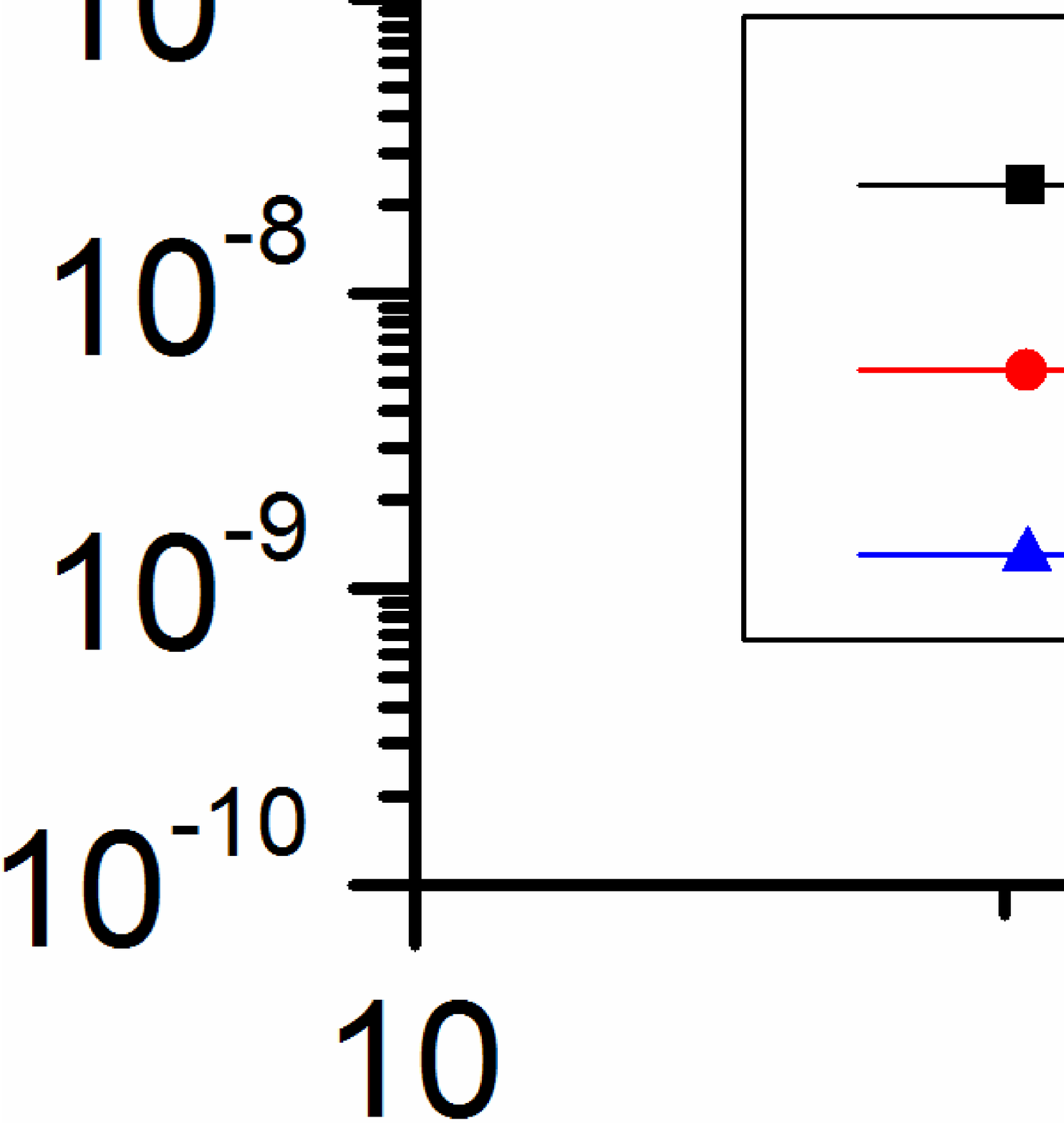}
		\caption{}
	\label{Fig5}
\end{figure*}

\begin{figure*}[htbp]
	\centering
		\includegraphics[width=0.50\textwidth]{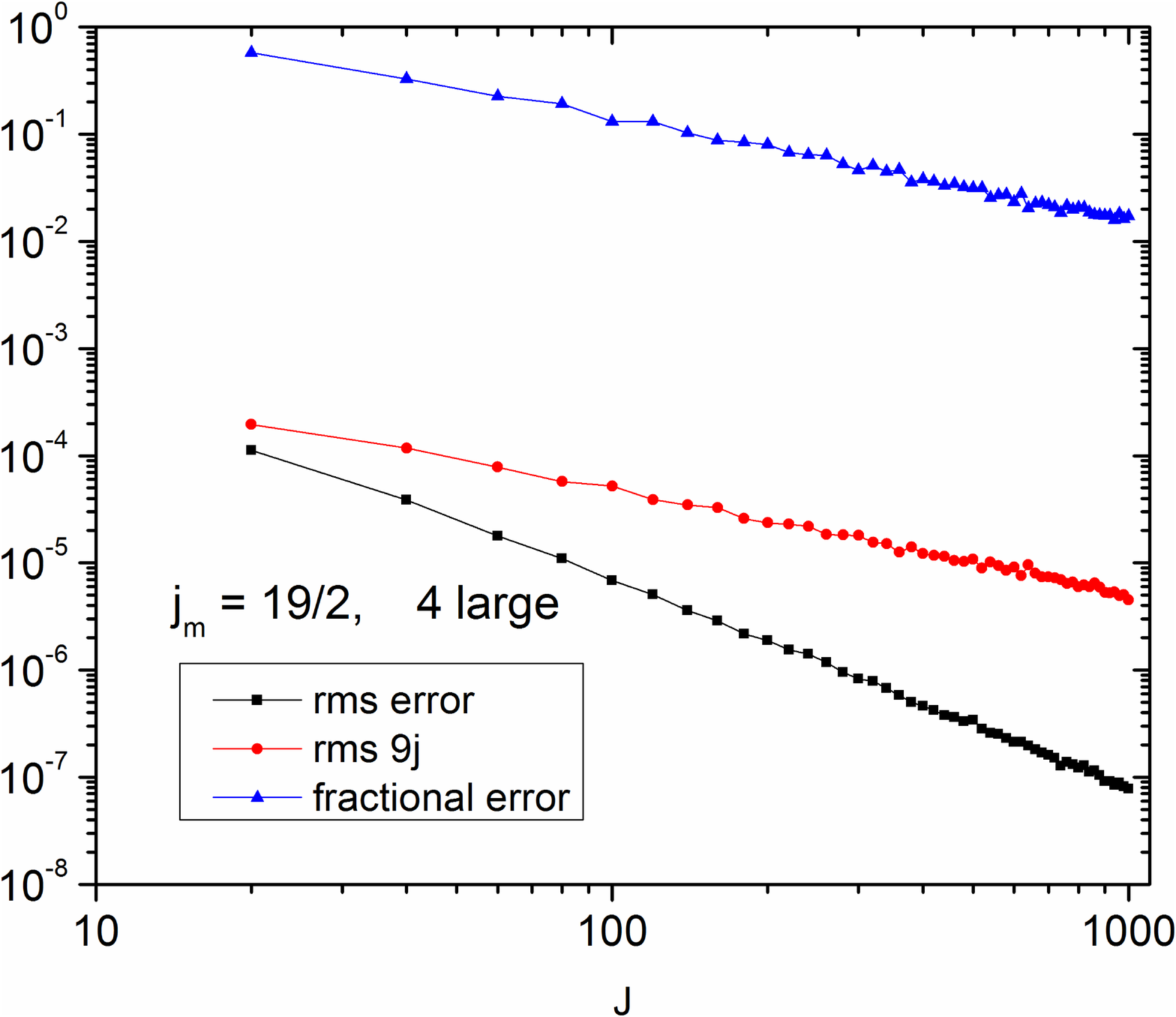}
		\caption{}
	\label{Fig6}
\end{figure*}

\begin{figure*}[htbp]
	\centering
		\includegraphics[width=0.50\textwidth]{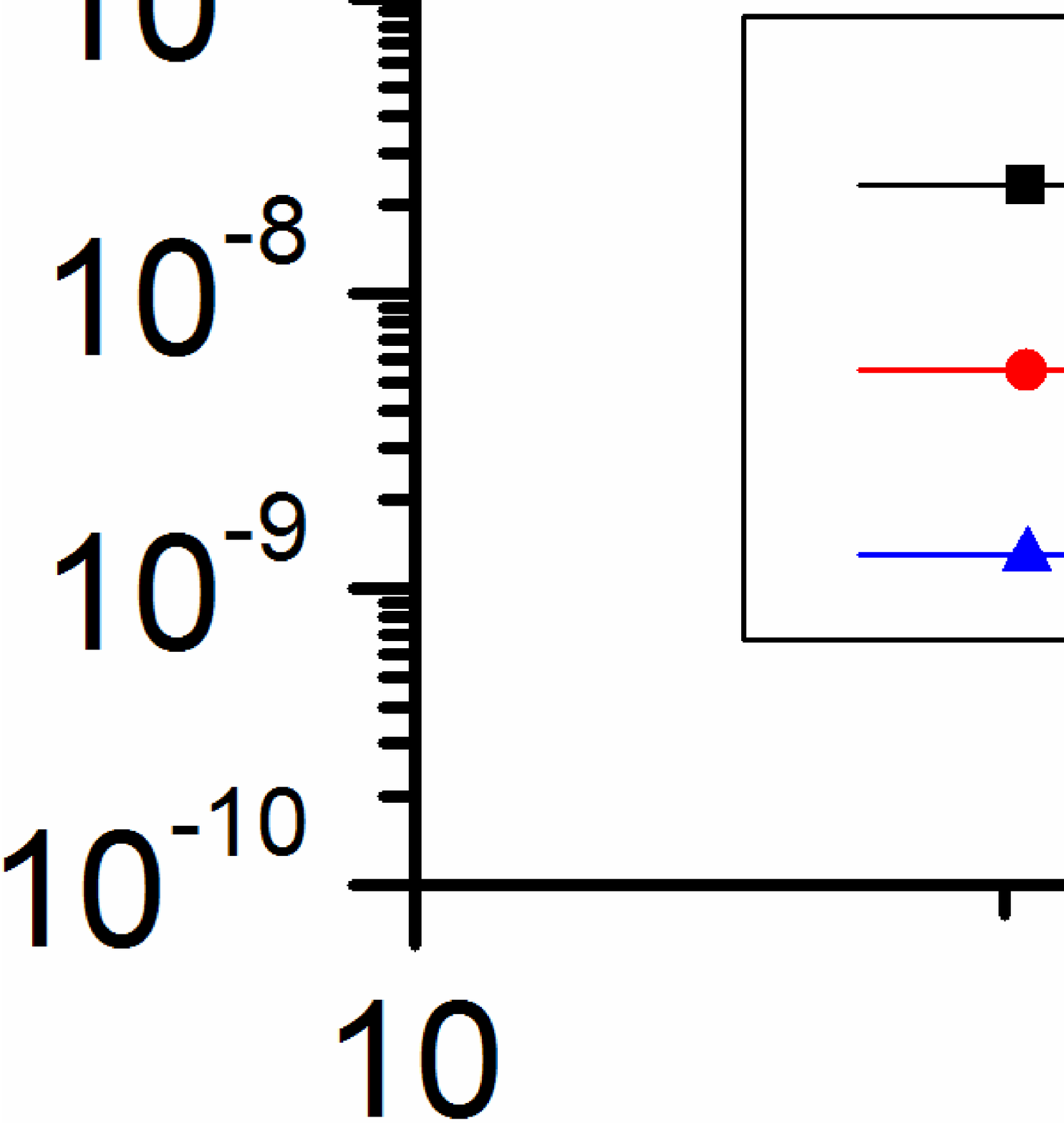}
		\caption{}
	\label{Fig7}
\end{figure*}

\begin{figure*}[htbp]
	\centering
		\includegraphics[width=0.50\textwidth]{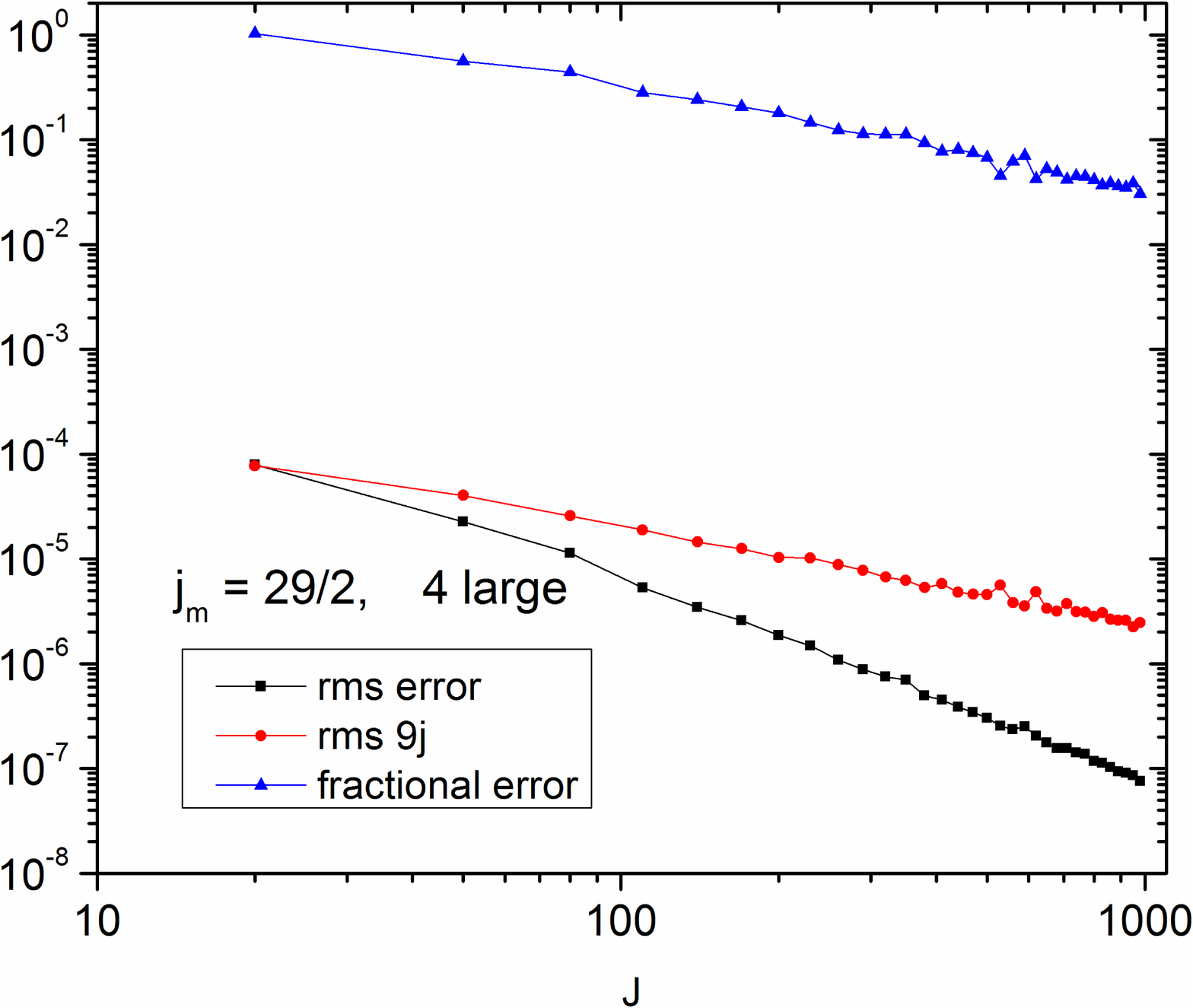}
		\caption{}
	\label{Fig8}
\end{figure*}

\vfill
\newpage

 %%%%%%%%%%%%%%%%%%%%%%%%%%%%%
%%%%%%%%%%%%%%                   SECTION 4.2 (PROP. 5)
%%%%%%%%%%%%%%%%%%%%%%%%%%%%

\subsection{The general case}

Recall from section 3 that both $3nj$(I) and $3nj$(II) spin networks
possess a Hamiltonian circuit of length $2n$.
In this section we are going to show how  disentangling
--induced by letting the $2n$ spin variables on this
closed cycle to become large--
works for any $3nj$ diagram of type I by applying 
recursively an asymptotic  version of the insertion 
operation $\mathfrak{I}_{\bowtie\,}$
(a similar procedure might be set up for $3nj$(II)).

In order to make the notation used in section 4.1
compatible with the general case, let us rewrite the
$9j$(I) accordingly in terms of the labels $\{j,k,l\}$
(see (\ref{RFtypeI}) and the diagram
of the $3nj$(I) at the beginning of section
3.2). Consider at the same time the limit in which
the Hamiltonian cycle (the external circle of the cartwheel
diagram with 3 internal edges) is labeled by
large spin values,  denoted by capital letters
\begin{equation}\label{9jJKl}
\begin{Bmatrix}
 j_1 & \, & j_2 & \,& j_3 & \, \\
 \, & l_1 & \, & l_2 &\, & l_3 \\
 k_1 & \, & k_2 &\, & k_3  & \, 
 \end{Bmatrix}
 \xrightarrow{j_1,j_2,j_3,k_1,k_2,k_3\gg 1}
 \begin{Bmatrix}
 J_1 & \, & J_2 & \,& J_3 & \, \\
 \, & l_1 & \, & l_2 &\, & l_3 \\
 K_1 & \, & K_2 &\, & K_3  & \, 
 \end{Bmatrix}
\end{equation}
To such a $(6,3)$ expansion of the $9j$ 
($6$ large, $3$ small entries) we can apply
the general twisted insertion operation defined in
Prop. 4 of section 4.2 to get the $12j$(I). 
In the present case, without imposing any restriction
on the entries, it would read
\begin{equation}\label{TwiLaIn}
\mathfrak{I}_{\bowtie\,}\;\leftrightarrow\;
\sum_x
\begin{Bmatrix}
 j_1 & k_{3} & x \\
 l_{3} & l_4 & k_4 
 \end{Bmatrix}
 \begin{Bmatrix}
k_1 & j_{3} & x \\
 l_{3} & l_4 & j_4 
 \end{Bmatrix}
\end{equation}
However, since such a combination has to be "coupled" with the $(6,3)$
$9j$ (in which an $x$ takes formally the place of the
"small" $l_3$ in (\ref{9jJKl})) we see that the entries
$j_1,k_3,k_1,j_3$  are necessarily "large"
because they lie on the preexisting Hamiltonian great circle,
while $l_3$ stays small. Then we are left with the residual problem
of determining the order of magnitude
of the new entries $j_4,k_4,l_4$ (recall that $x$ is small).
This is achieved by checking the triads of the two $6j$ symbols in (\ref{TwiLaIn}) 
to see which possibilities we have for these three labels
in view of the admissibile configurations for the $3j$ coefficient
discussed in the introduction to this section.
\begin{itemize}
\item The triads of
$\left\{\begin{smallmatrix}
 J_1 & K_{3} & x \\
 l_{3} & l_4 & k_4 
\end{smallmatrix}
\right \} $ are 
\vskip 6pt
$(J_1 \, K_3 \, x)$ (2 large, 1 small) and as such it is admissible;\\
$(l_3 \, l_4  \,x)$ (2 small) and thus $l_4$ must be small;\\
$(J_1  \, l_4 \,  k_4)$ (1 large, 1 small) and then  $k_4$ must be large
and renamed $K_4$;\\
$(l_3 \, K_3  \,K_4)$ (2 large, 1 small) and then admissible.

\item The triads of
$\left\{\begin{smallmatrix}
 K_1 & J_{3} & x \\
 l_{3} & l_4 & j_4 
\end{smallmatrix}
\right \} $ are
\vskip 6pt
$(K_1 \, J_3 \, x)$ (2 large, 1 small) and as such it is admissible;\\
$(l_3 \, l_4  \,x)$ is the same triad considered before;\\
$(K_1  \, l_4 \,  j_4)$ (1 large, 1 small) and then  $j_4$ must be large
and renamed $J_4$;\\
$(l_3 \, J_3  \,J_4)$ (2 large, 1 small) and then admissible.
\end{itemize}

Since the above remarks  clearly hold true for any
insertion operation which acts on a preexisting 
[$2(n-1)$ large, $(n-1)$ small] $3(n-1)j$(I) diagram,
we have established the following
\vskip 6pt
\noindent {\bf Proposition 5}.\\
{\em There exists (up to symmetry) one kind
of twisted "asymptotic" insertion operation, denoted by
 $\mathfrak{I}^{\text{as}}_{\bowtie}\,$, that generates
 a unique Hamiltonian disentangled $(2n,n)$ 
$3nj$(I) diagram  from the similar $[2(n-1),(n-1)]$  
$3(n-1)j$(I) for
any $n\geq 3$.} 
\vskip 6pt
Its general (graphical and algebraic) expression reads 
 \[
\xy
0*{\mathfrak{I}^{\text{as}}_{\bowtie}};
<1cm,0cm>*{\leftrightarrow};
<2cm,-0.5cm>*{\bullet}="A";
<3.5cm,-0.5cm>*{\bullet}="B";
<3.5cm,0.5cm>*{\bullet}="C";
<2cm,0.5cm>*{\bullet}="D";
<1.5cm,-1cm>*{}="A'";
<4cm,-1cm>*{}="B'";
<4cm,1cm>*{}="C'";
<1.5cm,1cm>*{}="D'";
"A";"A'" **@{:} ?(0.3)*!/_4mm/{K_{n-1}};
"B'";"B" **@{:} ?(0.3)*!/_2mm/{K_1};
"C";"C'" **@{:} ?(0.7)*!/_4mm/{J_{n-1}};
"D'";"D" **@{:} ?(0.3)*!/_3mm/{J_1};
"A";"D" **@{:} ?(0.5)*!/_3mm/{K_n};
"C";"B" **@{:} ?(0.5)*!/_2mm/{J_n};
"C";"A" **@{-} ?(0.7)*!/_3mm/{l_{n-1}};
"D";"B" **@{-} ?(0.2)*!/_2mm/{l_n};
<8cm,0cm>*{\leftrightarrow
\, \sum_x
\begin{Bmatrix}
 J_1 & K_{n-1} & x \\
 l_{n-1} & l_n & K_n 
 \end{Bmatrix}
 \begin{Bmatrix}
J_1 & J_{n-1} & x \\
 l_{n-1} & l_n & J_n 
 \end{Bmatrix}
};
\endxy
\]
where edges with large labels are drawn as
dotted lines.

Thus we can easily infer
that, on the basis of the $(6,3)$ expansion of the $9j$ given in   (\ref{9j63}),
the analytical expression of any 
Hamiltonian asymptotic expansion of a  
$(2n,n)$ 
$3nj$(I) diagram contain the product (no sum)
of $(n-1)$ Wigner $\mathbf{d}$--functions
whose principal quantum numbers may be chosen
among the $n$ spin variables $\{l_1,l_2, \dots,
l_n\}$
(a similar result can be established for $3nj$(II) diagrams).

%%%%%%%%%%%%%%%
%%%%%%%%%%%%%%SECTION 5
%%%%%%%%%%%%%%%%%

\section{Conclusions and outlook}

We have developed in this paper 
a recursive  procedure for the generation of $3nj$
from $3(n-1)j$ diagrams  of types I and II
as well as an original approach to the classification
of their asymptotic disentangling.

There are a number of open questions that
deserve further investigations.\\
The recursive generation
of all other types of $3nj$ diagrams would be a major achievement,
as pointed out already in \cite{BiLo9} (topic 12). We argue that
it would be possible (at least in principle) to complete the enumeration
of diagrams of type III, IV, V,... for any  $n > 4$
starting from either type I or II and applying suitably chosen sequences
of the two kind of insertion operators defined in section 3.
However it should be clear that the enumeration will blow up as $n$
increases and the hardest task will be to rule out isomorphic 
configurations (see footnote 2 and \cite{Belgi}).

For what concerns semiclassical limits and
asymptotic disentangling, we are currently
performing numerical experiments on the five types
of $15j$ coefficients to test the relative magnitudes and associated
probabilities ({\em cfr.} \cite{AmRo} for an early attempt
of comparison between type I and II).
We have restricted most of the treatment in section 4 to Hamiltonian
circuits, but it can be conjectured that the described features of disentangling
will occur for other circuits in general spin networks.
It would be also interesting to analyze the different asymptotic regimes
in connection with the symmetry properties of the different 
types of coefficients.\\
An improvement of 
multi--variable recursion relations and
related partial differential equations 
(see the $9j$ case discussed in section 2.2)
would greatly help in providing 
further insights in the classification
of the associated orthogonal polynomials 
of hypergeometric type.

The results obtained here, as well as possible
new improvements along the lines outlined above, represent
a prerequisite for extending to $3nj$ the
analysis based on  sophisticated (quantum and semiclassical) 
geometric and analytical  techniques
introduced in \cite{AqHaLi1,AqHaLi2}, see also  
\cite{LiYu}.

\section*{Acknowledgments}

 We are very pleased to co--author this paper with Professor Vincenzo Aquilanti in
this issue in honor of his seventieth birthday.  We join in celebrating his
remarkable career in theoretical and experimental chemistry.  We have been Enzo's
friends for many years, and we look forward to much more good times and excellent
science (R.A. and A.M.).

\end{document}